\documentclass[a4paper,onecolumn,11pt,accepted=2026-08-29]{quantumarticle}
\pdfoutput=1
\usepackage[utf8]{inputenc}
\usepackage[english]{babel}
\usepackage[T1]{fontenc}
\usepackage{amsmath}
\usepackage{hyperref}
\usepackage[numbers,sort&compress]{natbib}
\usepackage{tikz}
\usepackage{lipsum}

\usepackage{ragged2e}
\usepackage{subfig}
\usepackage{stmaryrd}
\usetikzlibrary{quantikz}
\usepackage{amsmath, esint, amsthm, physics, amsfonts, amssymb} 
\usepackage{graphicx, float} 
\usepackage{multirow}
\newcommand{\coolrightbrace}[2]{\mathclap{\left.\vphantom{\begin{matrix} #1 \end{matrix}}\right\}}\quad#2}

\DeclareMathSymbol{\shortminus}{\mathbin}{AMSa}{"39}

\begin{document}

\title{Solving the Nonlinear Vlasov Equation on a Quantum Computer}

\author{Tamás Vaszary}
\affiliation{Department of Physics, University of Warwick, Coventry CV4 7AL, United Kingdom}
\affiliation{Mathematical Institute, University of Oxford, Oxford OX2 6GG, United Kingdom}
\affiliation{Department of Physics, Imperial College London, London SW7 2AZ, United Kingdom}
\orcid{0000-0001-7811-222X}
\email{tamas.taszary@maths.ox.ac.uk}

\author{Animesh Datta}
\affiliation{Department of Physics, University of Warwick, Coventry CV4 7AL, United Kingdom}
\orcid{0000-0003-4021-4655}
\email{animesh.datta@warwick.ac.uk}

\author{Tom Goffrey}
\affiliation{Department of Physics, University of Warwick, Coventry CV4 7AL, United Kingdom}
\orcid{0000-0003-0784-1294}

\author{Brian Appelbe}
\affiliation{Department of Physics, Imperial College London, London SW7 2AZ, United Kingdom}
\orcid{0000-0003-4781-7010}

\maketitle

\begin{abstract}

The practical applicability of a recent Carleman-linearization-based quantum algorithm for solving ordinary differential equations (ODEs) with quadratic nonlinearities is investigated for the nonlinear electrostatic Vlasov equation with Krook-type collision operators. The equation is discretized on a $(1+1)$-dimensional phase-space grid and mapped onto the input of the quantum algorithm. Upper bounds for the query and gate complexities are derived in the limit of large grid sizes and found to be polynomially larger than the time complexity of the corresponding classical algorithms, primarily due to the dimension, sparsity, and norm of the Carleman-linearized evolution matrix. The convergence criteria are shown to impose severe restrictions on physically relevant plasma applications, requiring dissipation levels far exceeding those provided by the Krook operator.
\end{abstract}

\tableofcontents

\section{Introduction}
The ability to effectively simulate and predict plasma behaviour is key to our understanding of a rich variety of physical phenomena, such as high-energy astrophysics, as well as magnetic- and inertial confinement fusion. However, the capacities of current classical supercomputers often prove to be insufficient for large-scale plasma simulations because of the inherently multi-scale nature of the problem, where fluid and kinetic scales are coupled across spatial and temporal domains. Alternatives to classical computation are thus desirable. One of the candidates is quantum computation, which offers more efficient algorithms for certain problems in terms of computational complexity \cite{shor,grover}. Consequently, quantum computation is beginning to attract attention in the plasma physics community. 
Recent reviews \cite{Dodin, Joseph,living} identify some of the challenges of mapping plasma problems onto the architecture of quantum computers. 
The most notable difficulty that arises in the construction of such mappings is the fact that while plasma models are generally nonlinear, quantum computers are naturally suited to linear problems only. So far, this difficulty has often been mitigated by focusing on plasma problems in the linearized regime \cite{Novikau,plasmaquant1,plasmaquant2,novikau2024encoding,wave,neutrino}. However, the practicality of this approach is limited as most plasma phenomena of scientific relevance, such as turbulence, shock waves and wave-particle interactions are strongly nonlinear \cite{plasma2}.

Quantum approaches to nonlinear dynamics have gained more attention in the field of computational fluid mechanics. For the Navier-Stokes equation, a specific method was introduced in Refs. \cite{gaitan1,gaitan2}, with a broader discussion provided in Ref. \cite{navier_stokes}. The convergence of the Carleman linearization for the Navier-Stokes equations was examined in Ref. \cite{linearization_efficiency}, following a methodology similar to the one we describe later. Quantum solutions of fluid dynamics given by the Burgers' and Navier-Stokes equations have been studied using the lattice-Boltzmann formulation as well as  Carleman linearization, a technique we revisit in this work \cite{lattice_boltzmann_1,lattice_boltzmann_5,lattice_boltzmann_3,demirdjian2025efficientdecompositioncarlemanlinearized, liu,Costa_2025}.

The Vlasov equation, which we focus on in this paper, is particularly important for the simulation of nonlinear plasma phenomena. Unlike macroscopic fluid-based descriptions, the Vlasov equation emerges from kinetic theory and provides a significantly more detailed way of capturing plasma dynamics. Classical algorithms that solve the Vlasov equation (coupled to Maxwell's equations) include grid-based and operator-splitting methods, as well as particle-in-cell simulations \cite{vlasov_solver_1,vlasov_solver_2,vlasov_solver_3,vlasov_solver_4}.
The Vlasov-Maxwell system is a family of coupled nonlinear partial differential equations (PDEs).

The first quantum algorithm to solve nonlinear differential equations had a runtime scaling exponentially with simulation time \cite{sarah}. This was later improved to quadratic \cite{lloyd} but without proving the correctness of their approach.
More recently, the mapping of nonlinear classical dynamics to linear quantum dynamics has been investigated \cite{embedding,koopman7,koopman6,Julien2,Julien}, however these typically do not provide generally applicable methods to perform the mappings of arbitrary (polynomially) nonlinear dynamics.

Another direction aimed at procedural implementation is based on embedding methods such as, homotopy perturbation \cite{homotopy} or Carleman linearization \cite{liu,liu2,surana2023efficient,krovi,variational_Carleman,Brustle2025quantumclassical, Costa_2025,jennings2025quantumalgorithmsgeneralnonlinear}. These approaches rely on an already well-developed understanding of quantum algorithms for linear ODEs~\cite{an2023theoryquantumdifferentialequation, berry1,berry2}. They lead to quantum algorithms with favourable query- and gate-complexity scaling with respect to simulation time, the number of dynamical variables, and the allowed solution error. This scaling and the procedural applicability comes at the cost of two limitations.
Firstly, the linear part of the evolution is required to be dissipative, 
meaning that the norm of the solution vector decreases in time. Secondly, the ratio of the combined strength of nonlinear and inhomogeneous effects to that of linear dissipation is required to be small. 
This is typically formulated as $R<1$ for some so-called convergence parameter $R$ quantifying the mentioned ratio. For a discussion about relaxing these requirements, we direct the reader to Refs.~\cite{no_dissipative,jennings2025quantumalgorithmsgeneralnonlinear,wang2026quantumalgorithmsnonlineardifferential,Costa_2025}. 

In this paper, we provide a quantum algorithm for solving the nonlinear Vlasov equation in the non-relativistic, electrostatic limit (Eq.~\eqref{eq:original_vlasov}).
It is based on a quantum algorithm due to Krovi~\cite{krovi}. 
For simplicity, we focus on one position and one velocity dimension.
The two-dimensional continuous phase space is discretized and the Vlasov equation is mapped to a finite difference equation, compatible with the mathematical framework of Ref.~\cite{krovi}. The electrostatic interaction is captured by Gauss's law. This results in a family of ordinary differential equations (ODEs) with a quadratic nonlinearity.
We embed these into an infinitely large family of linear ODEs through Carleman linearization \cite{carleman0}. After truncation to a finite family, it is evolved by a Taylor series-based high-order time integrator. Due to the time discretization, the dynamics are represented as a matrix inversion problem, which we solved with a Quantum Linear Solver Algorithm (QLSA) \cite{dervovic2018quantumlinearsystemsalgorithms,harrow}. Furthermore, we show that when the Vlasov equation is coupled to Gauss's law, the constraints mentioned in the previous paragraph are satisfied for certain regimes of plasma parameters as long as collisions make the system dissipative enough. On the other hand, when coupling it to the otherwise physically equivalent Ampere's law, we find these convergence criteria to be unsatisfiable regardless of the values of the plasma parameters. This is an illustration of different numerical routes to solving the same physical problem having different regimes of validity.

The primary contribution of this work is not merely the construction of a mapping between the Vlasov equation and the quantum algorithm of Ref.~\cite{krovi}, but the assessment of whether that algorithm can operate in physically relevant plasma regimes. We show that the convergence conditions required by the Carleman linearization impose stringent dissipation requirements that are incompatible with typical plasma parameters. Consequently, our analysis provides a quantitative characterization of the limitations of the current quantum algorithmic framework for nonlinear plasma simulations.

Our work reaches the following technical conclusions. First, we find that the dissipation mechanism of the plasma (originating from collisions and modeled by a Krook-type collision operator) directly translates into the dissipation of the nonlinear ODE, which is required for the convergence of Carleman linearization. We also find that physically it is far too weak to lead to $R$ parameter values less than unity for realistic grid-sizes. This is because its numerator contains $\|F^{(2)}\|$, the norm of the matrix encoding the nonlinearity. It contains a double integral in our case that computes the electric field from Gauss's law, and causes $\|F^{(2)}\|$ to grow with grid-size. Subsection \ref{sec:connection_to_plasma} gives an explicit upper bound on the number of velocity grid points in terms of plasma parameters, such as temperature and the volume of the phase space. We use typical numerical parameters from inertial confinement fusion and interstellar plasma to illustrate the severity of this restriction. Since the original preprint of this work, related developments building on and extending Carleman-based approaches to nonlinear Vlasov systems have appeared \cite{jennings2025quantumalgorithmsgeneralnonlinear,christensen2026improvedconvergencecarlemanembeddedquantum,berntson2026endtoendquantumalgorithmweakly}; see Section~\ref{section:conclusion} for a discussion of their implications for our quantitative conclusions.

Second, we find that the query- and gate complexities are polynomially larger than the time complexity of the simple classical implementation of the same finite difference scheme, with respect to error and simulation time, in the asymptotic limit of large grid sizes. The main contributors to these complexities are the number of required Carleman linearization steps enlarging the dimension of the linearized system, the norm of the matrix encoding that system, and finally the sparsity. The latter scales linearly with system size due to Gauss's law containing a double integral over phase space. The upper bounds for the final complexities are presented in Eqs. (\ref{eq:T_eps_querycomplexity_final}-\ref{eq:T_eps_gatecomplexity_final}).

\begin{figure}[h!]
    \centering
\includegraphics[width=6 in]{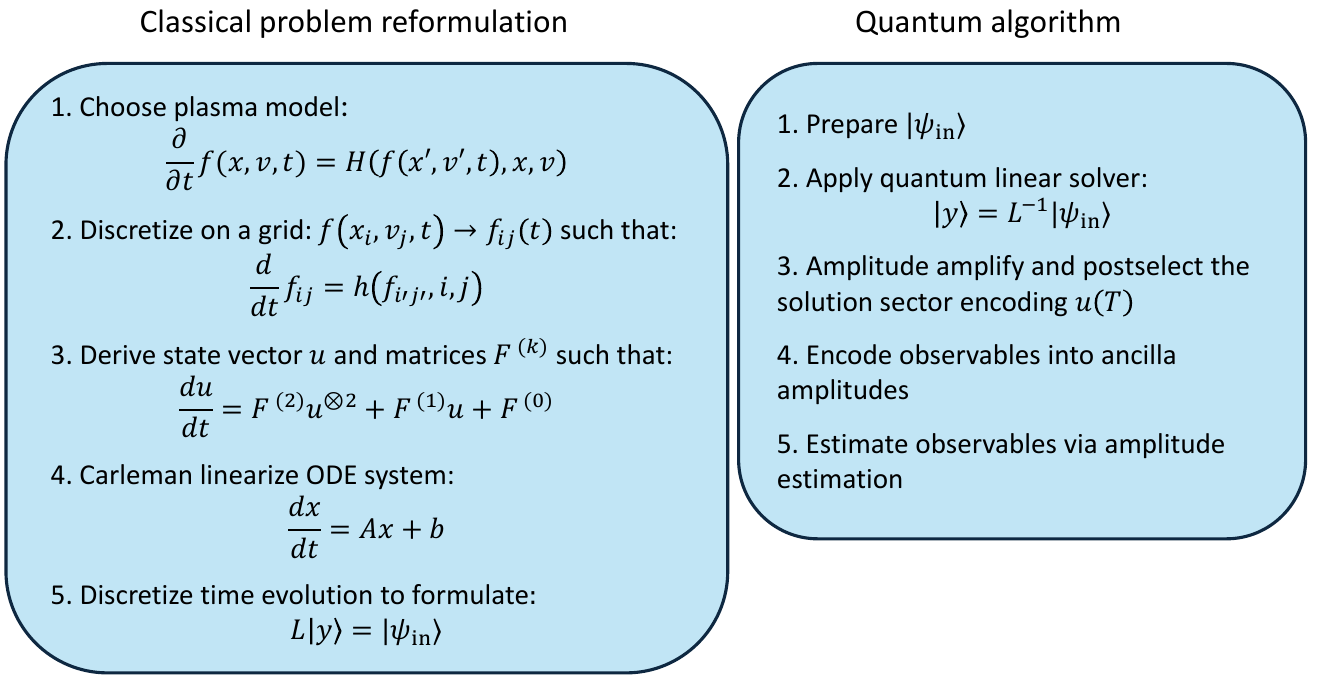}
\caption{A schematic of the procedure used in this paper. The left panel shows the classical reformulation pipeline: the plasma model is discretized, rewritten as a finite-dimensional nonlinear ODE system, Carleman linearized, and converted into the linear system $L|y\rangle = |\psi_{\rm in}\rangle$. This linear system is then passed to the quantum algorithm shown in the right panel. Steps 1--3 summarize the algorithmic pipeline in Ref. \cite{krovi}: the input state $|\psi_{\text{in}}\rangle$ is prepared, the quantum linear solver is applied to obtain $|y\rangle=L^{-1}|\psi_{\text{in}}\rangle$, and the solution sector encoding $u(T)$ is amplified and postselected. Steps 4--5 illustrate, at a high level, a possible observable-extraction procedure for physically relevant quadratic functionals. Note that in Ref. \cite{krovi}, the problem statement is the output of Step 3 on the left panel.}
\label{fig:flowchart}
\end{figure}

Our procedure is summarized in Fig. \ref{fig:flowchart}. It presents the first step on the path to solving general plasma physics problems on quantum computers.
The approach taken builds on the preliminary results developed in previous work \cite{dissertation}. 

This paper is structured as follows. 
For clarity, we first establish the notation used in the paper in Table \ref{tab:notation}. 
In Section \ref{vlasov_section}, we present the finite difference equation to be solved. 
In Section \ref{section:quantum_algo}, we summarize the relevant parts of the quantum algorithm from Ref.~\cite{krovi} for completeness. 
In Section \ref{sec:mapping}, we construct the structure of the mapping between the Vlasov equation coupled to Gauss's law and the input of Ref.~\cite{krovi}. 
In Section \ref{convergence}, we investigate the behaviour of the convergence criteria specifically for the finite difference equations in Section \ref{sec:gauss_law_finite_diff}. In Section \ref{section:errors}, we calculate the propagation of the phase space discretization error through the algorithm. 
In Section \ref{complexity_section}, we present the upper bounds of the complexities of our quantum algorithm. In Section \ref{sec:ampere} we analyse how coupling to Ampere's law instead of Gauss's law affects the convergence of Carleman linearization. We finish by discussing our results in Section \ref{section:discussion} and concluding in Section \ref{section:conclusion}.

\begin{table}[h!]
\centering
\begin{tabular}{|c|c|||c|c|} 
 \hline
 \textbf{Symbol} & \textbf{Meaning} & \textbf{Symbol} & \textbf{Meaning}\phantom{\Big|} \\ [0.4ex] 
 \hline\hline
 \phantom{\Big|} $v_n$ &  $n$-th entry of vector $v$ &  $\kappa(A)$  & condition number of matrix $A$ \\ 
 \hline
 \phantom{\Big|} $\norm{v}$ &  $l_2$ norm of vector $v$ & $\text{vec}(A)$ & vectorization of matrix $A$ \\  
 \hline
 \phantom{\Big|} $A_{ij}$ & $ij$-th entry of matrix $A$ & $\mathbb{I}$ & identity matrix\\ 
 \hline
 \phantom{\Big|} $\norm{A}_k$ &  induced $l_k$ norm of matrix $A$ & $\lceil x\rceil$ & ceiling of number $x$\\
 \hline
 \phantom{\Big|} $\norm{A}=\norm{A}_2$ &  spectral norm of matrix $A$ & $x\sslash y$ & $ x-y\left(\left\lceil x/y\right\rceil-1\right)$\\
 \hline
 \phantom{\Big|} $\norm{A}_F$ &  Frobenius norm of matrix $A$ & $\text{erf}(x)$ &  $(2/\sqrt{\pi})\int_0^x \exp(-t^2)dt$ \\
 \hline
 \phantom{\Big|} $\mu(A)$ &  log-norm of matrix $A$ & $\delta_{ij}$ & Kronecker delta\\
 \hline
  \phantom{\Big|} $\alpha(A)$& spectral abscissa of matrix $A$ & $\oplus$ & direct sum \\ 
 \hline
\end{tabular}
\caption{Summary of the notation used in our paper. More details are in Appendix \ref{appendix:maths}}
\label{tab:notation}
\end{table}

In Appendix \ref{appendix:maths} we define the mathematical expressions used in our paper. Next, in Appendix \ref{appendix:vectorization_of_u_gauss}, we carry out the vectorization of the dependent variables, leading to the vectors $u$ and $u^{\otimes 2}$ that appear in Step 3 of the left panel of Fig. \ref{fig:flowchart}. In Appendix \ref{appendix:entries} we state the entries of the corresponding $F^{(k)}$ matrices. In Appendices  \ref{appendix:norm_derivations} and \ref{appendix:quantities_for_complexity} we derive quantities from these objects needed for the complexity. Finally, in Appendix \ref{appendix:classical_error_complexity} we perform the error and complexity analysis of the classical solution of the problem.

\section{The Vlasov equation}\label{vlasov_section}
\subsection{The model}
The Vlasov equation describes the time evolution of the phase space distribution function $f(x,v,t)\in \mathbb{R}_{\geq 0}$, where $x\in \mathbb{R} $ is position and $v\in \mathbb{R}$ is velocity. 
We restrict ourselves to one space and one velocity dimension and apply it to a plasma with one dynamical species, electrons, with charge $-q<0$ and mass $m_e$. 
An additional fixed, neutralising and stationary ion background species is denoted by $f^{\text{bg}}(x,v)=f^{\text{bg}}(v)$, which is taken to be uniform in space. This choice does not affect the convergence of our quantum algorithm. Also note that we use SI units throughout this paper. Then the Vlasov equation for $f(x,v,t)$ in the electrostatic limit reads
\begin{align}
\frac{\partial f}{\partial t} +v\frac{\partial f}{\partial x}-\frac{qE}{m_e}\frac{\partial f}{\partial v}  =C[f],\label{eq:original_vlasov}
\end{align}
where $C[f]$ is a collision operator that relaxes $f$ into a target distribution, conventionally chosen to be a Maxwellian. $E(x,t)$ is the electric field generated by the total (electron and ion) charge distribution. Combining with Gauss's law the full system reads
\begin{align}
\frac{\partial f}{\partial t} =-v\frac{\partial f}{\partial x}+\frac{q^2}{m_e\varepsilon_0}\frac{\partial f}{\partial v}\int^x dx'\int dv' \left( f^{\text{bg}}-f \right)  +C[f].\label{eq:onespecisesonedimension}
\end{align}
However, we can treat the field as a dynamical variable and evolve it using Ampere's law. Then the resulting system is 
\begin{align}
\begin{split}
\frac{\partial f}{\partial t}&= -v\frac{\partial f}{\partial x}+\frac{q}{m_e}E\frac{\partial f}{\partial v}  +C[f],\\
\frac{\partial E}{\partial t}&=\frac{q}{\varepsilon_0}\int v\left(f-f^{\text{bg}}\right)dv.\label{eq:ampereevolution}
\end{split}
\end{align}

We take periodic boundary conditions (BCs) in position with periodicity $x_{\max}$, as well as fixed ones in velocity such that $f(x,v,t)$ goes to zero outside $|v|=v_{\max} $, for some $v_{\max} $:
\begin{subequations}
\begin{align}
f(0,v,t)&=f(x_{\max},v,t),\quad &\forall\:v,t,\label{eq:bcsx}\\
f(x,v,t)&=0,&\forall \: |v|>v_{\max} \text{ and } \forall\: x,t,\label{eq:bcsv}\\
E(0,t)&=E(x_{\max},t)=0,& \forall\:\phantom{x,} t.\label{eq:bcsE}
\end{align}
\end{subequations}
Note that these are chosen for mathematical simplicity and do not affect the conclusion of this work. Finally, we use a generalized version of the Krook collision operator \cite{krook1}, which is given by
\begin{align}
C\left[f\right]\equiv\nu(v)\left(f^{M}-f\right),\label{eq:krook}
\end{align}
where $f^{M}=f^{M}(v) \sim \exp(-b v^2)$ is a Maxwellian distribution, which is a Gaussian in one dimension, and $\nu(v)$ is a velocity dependent collision frequency. The Krook collision operator pushes $f$ towards $f^{M}$, the target Maxwellian, over timescale $1/\nu(v)$. 
Physically, the parameter $b$ can be expressed as $b=m_e/(2k_B\mathcal{T})$, where $k_B$ is the Boltzmann constant and $\mathcal{T}$ is the temperature. We take $\nu(v)$ to be in the general form
\begin{align}
\nu(v)\equiv \nu_0+h(v),\label{eq:defofnu}
\end{align}
where $\nu_0$ is a base-line collision frequency and $h(v)=h(-v)>0$ is a small (positive) variation on top of it. This form is chosen such that $\min_{v} \nu(v)=\nu_0>0$ and
\begin{align}
\max_{|v|\leq v_{\max}}h(v) \ll \nu_0,\label{eq:h_small}
\end{align}
which will be used in proving the convergence of the quantum algorithm.

The velocity dependence of $\nu(v)$ in the Krook model was proposed in Ref. \cite{krook2} and further investigated in Refs. \cite{krook3,krook4}. Their conclusion is that by enforcing certain conditions on the function $\nu(v)$ (which we do not explicitly do here), the Krook operator can mimic the behaviour of the physically more meaningful Fokker–Planck operator \cite{fokker} despite being more simple. Furthermore, this can be done while maintaining some desired properties of the system such as the conservation of particle number, total momentum and energy.

\subsection{Discretization of phase space}\label{sec:coord_discretization}
We discretize the distribution function $f(x,v,t)$ on an $N=N_x \times N_v $ grid, leading to
\begin{align}
f \equiv
\begin{pmatrix}
f_{1,1} &\cdots &f_{1,N_v} \\
\vdots &\ddots &\vdots\\
f_{N_x,1} &\cdots &f_{N_x,N_v}\\[0.1cm]
\end{pmatrix},\label{eq:distributionmatrix}
\end{align}
where $f_{i,j}  \equiv f(x_i,v_j,t)$ and the grid is spanned by $N_x$ independent spatial coordinates indexed by $i$ and $N_v$ velocity coordinates indexed by $j$. We define
\begin{align}
\begin{split}
&x_i\equiv(i-1)\Delta x,\quad i=1,\dots,N_x,\\
&\text{such that} \quad x_1=0,\:x_{N_x}=x_{\max}-\Delta x\quad \quad \text{and} \quad \quad  x_{\max} \equiv N_x\Delta x,\label{eq:xcoordinate}
\end{split}
\end{align}
and
\begin{align}
\begin{split}
&v_j \equiv -v_{\max}+(j-1)\Delta v,\quad j=1,\dots, N_v,\\
&\text{such that}\quad v_1=-v_{\max},\:v_{N_v}=v_{\max} \quad \quad \text{and} \quad \quad  v_{\max}\equiv\frac{(N_v-1)\Delta v}{2}.\label{eq:vcoordinate}\\
\end{split}
\end{align}

Furthermore, we choose to approximate the differential operators $\partial/\partial x,\: \partial/ \partial v$ with second order central derivatives and the integrals with the trapezoidal rule, as given in Appendix \ref{discrete_calculus}. These introduce an error of $\mathcal{O}(\Delta x^2+\Delta v^2)$ on $d f_{ij}/d t$.

Note that we write $N = N_x N_v$ for the total number of phase-space grid points. The indices $N_x$ and $N_v$ are kept explicit in the mapping and convergence analysis, while the asymptotic complexity section assumes a fixed aspect ratio $N_x = \mathcal{O}(N_v)$, so that $N_x^2 = \mathcal{O}(N) = N_v^2$.

\subsection{Finite difference equations}\label{sec:gauss_law_finite_diff}
This paper, except for Section \ref{sec:ampere}, uses Gauss's law as a coupling between the Vlasov equation and Maxwell's equations; i.e., aims to solve Eq. \eqref{eq:onespecisesonedimension}. Applying the above discretization to that system results in the finite difference equations
\begin{subequations}
\begin{align}
\frac{d}{dt}f_{ij} =
& -\frac{q^2}{m_e\varepsilon_0} \left. \frac{\partial}{\partial v} \right|_{ij} f \cdot \iint^{x_i} f_{IJ}\, dx_I dv_J	
	&& \text{(quadratic)} \label{eq:1species_discretized_a} \\
& -v_j \left. \frac{\partial}{\partial x} \right|_{ij} f 
    + \frac{q^2}{m_e\varepsilon_0} \left. \frac{\partial}{\partial v} \right|_{ij} f \cdot \iint^{x_i} f^{\text{bg}}_{IJ}\, dx_I dv_J	
	&& \text{(linear)}\label{eq:1species_discretized_b} \\
& -\nu(v_j)f_{ij} 											
	&& \text{(linear)}\label{eq:1species_discretized_c} \\
& +\nu(v_j)f^{M}_j									
	&& \text{(inhomogeneous)} \label{eq:1species_discretized_d},
\end{align}
\end{subequations}
where the discretized target Maxwellian is
\begin{align}
f^{M}_j=\frac{\mathcal{N}}{2x_{\max} \Delta v}\Bigg[\sum_{J=1}^{N_v}\exp(-b v_J^2) \Bigg]^{-1}\exp\left(-bv_j^2\right).
\end{align}
The prefactor in the above equation ensures that $\iint fdxdv=\mathcal{N}$ under the trapezoidal rule for the integration on the whole grid ($0\leq x\leq x_{\max}=x_{N_x+1}$), where $\mathcal{N}$ is the number of electrons per unit area of the other $2$ position dimensions. Said otherwise, $\mathcal{N}=\int  n(x) dx$ where $n(x)$ is the 3 dimensional spatial density, taken to be a function of $x$. By assuming that the static background is uniform in space, the integral in Eq. \eqref{eq:1species_discretized_b} becomes
\begin{align}
\iint^{x_i}f^{\text{bg}}_{IJ}\:dx_Idv_J = \frac{i-1}{N_x}\mathcal{N}, \label{eq:integralofion}
\end{align}
where we take the same $\mathcal{N}$ as for electrons to ensure net charge neutrality.

\section{The quantum algorithm}\label{section:quantum_algo}
The quantum algorithm onto which the discretized Vlasov equation is mapped was recently proposed by Krovi \cite{krovi}, expanding on the foundational work of Berry et al. for solving linear differential equations \cite{berry1,berry2}. In \cite[Theorem 7]{krovi}, a linear ODE solver is presented, while in \cite[Theorem 8]{krovi}, it is applied to ODE systems with quadratic nonlinearities using Carleman linearization. We summarize the relevant parts of the quantum algorithm below.

\subsection{Problem statement}
The quantum algorithm in question solves an ODE system of the form
\begin{align}
\frac{du}{dt}=F^{(2)}u^{\otimes2}+F^{(1)}u+F^{(0)}, \quad \quad u(0)=u_{\text{in}},\label{eq:krovidef}
\end{align}
where the state vector $u\equiv[u_1,\dots,u_d]^\mathsf{T}\in \mathbb{R}^d$ is a $d$ dimensional column vector and $u^{\otimes2}\equiv u\otimes u=[u_1^2,u_1u_2,\dots,u_1u_d,u_2u_1,\dots,u_d u_{d-1},u_d^2]^\mathsf{T}\in \mathbb{R}^{d^2}$ is a $d^2$ dimensional column vector containing all the quadratic nonlinearities. $\mathsf{T}$ denotes the transpose. Each $u_k=u_k(t)$ is a function of time $t$ on the interval $t\in[0,T]$. 
The matrices $F^{(2)}\in\mathbb{R}^{d\times d^2}$ and $F^{(1)}\in\mathbb{R}^{d\times d}$, and the vector $F^{(0)}\in\mathbb{R}^{d}$ are all time independent and have sizes such that the matrix multiplications in Eq. \eqref{eq:krovidef} are well defined.

Other quantum algorithms based on procedural linear embedding apply to slightly different systems.
For instance, Refs.~\cite{homotopy} and \cite{liu2} require $F^{(0)}=0$ and Ref.~\cite{liu} requires $F^{(0)}=F^{(0)}(t)$, i.e., an explicit time dependence in the inhomogeneous term. This would correspond to an explicit time dependence in the target Maxwellian in the collision operator, i.e., $f^{M,(s)}(v)\to f^{M,(s)}(v,t)$. Finally, Ref.~\cite{surana2023efficient} is focused on higher order nonlinearities.

\subsection{Convergence criteria}\label{sec:convergence_citeria}
The quantum algorithm reviewed in Section \ref{subsec:qalgo} applies to the ODE system in Eq. \eqref{eq:krovidef} under two conditions.
Firstly, $F^{(1)}$ must have a negative log-norm:
\begin{align}
    \mu\left(F^{(1)}\right)<0,
\end{align}
which corresponds to the system being dissipative. Secondly, the convergence parameter $R$, given by
\begin{align}
R\equiv\frac{1}{|\mu(F^{(1)})|}\left(\norm{F^{(2)}}\norm{u_{\text{in}}} +\frac{\norm{F^{(0)}} }{ \norm{u_{\text{in}}}}\right),
\label{eq:R_def}
\end{align}
must satisfy $R<1$. Mathematically, these are both required to bound the error due to Carleman linearization \cite{forets2017explicit}. The norms are defined in Appendix \ref{appendix:norms}.

\subsection{The algorithm}
\label{subsec:qalgo}

\subsubsection{Input information}\label{rescaling}
The input consists of maps for the rescaled elements of each expression in Eq. \eqref{eq:krovidef}. Let us define the rescaling through the positive number $\gamma$ as
\begin{align}
u \to \Bar{u} \equiv \frac{u}{\gamma}.
\end{align}
This rescaling changes the $F^{(2)},\:F^{(0)}$ matrices as
\begin{align}
F^{(2)}\to \Bar{F}^{(2)} \equiv \gamma F^{(2)},\quad\quad F^{(0)} \to \Bar{F}^{(0)} \equiv \frac{F^{(0)}}{\gamma}.
\end{align}
Note that the rescaling does \emph{not} affect $F^{(1)}$ and $R$, and hence they will not carry the bar. The factor $\gamma$ must be chosen such that $\Bar{u}$ has no physical dimension and satisfies $\norm{\Bar{u}_{\text{in}}}<1$, as well as that
\begin{align}
\left |\mu\left(F^{(1)}\right)\right|>\norm{\Bar{F}^{(2)}}+\norm{\Bar{F}^{(0)}}.
\end{align}
These are required when bounding the error due to the Carleman linearization. 
Both these conditions can be met by~\cite{liu}
\begin{align}
\gamma \equiv \sqrt{\norm{u_{\text{in}}}r_+},\quad r_+ \equiv \frac{-\mu\left(F^{(1)}\right)+\sqrt{\mu\left(F^{(1)}\right)^2-4\norm{F^{(2)}}\norm{F^{(0)}}}}{2\norm{F^{(2)}}}\label{eq:gamma}.
\end{align}

The rescaled system is used henceforth. The corresponding oracle access model is described in Subsection~\ref{sec:oracles}, after the conceptual steps of the algorithm.

\subsubsection{Conceptual steps}
\begin{enumerate}
    \item The Carleman linearized system has a state vector $z$, given by
    \begin{align}
    z\equiv
    \left[\begin{array}{c}
    \Bar{u} \\
    \Bar{u}^{\otimes2} \\
    \vdots \\
    \Bar{u}^{\otimes N_C}
    \end{array}\right]=
    \left[\Bar{u}^\mathsf{T},\:{\Bar{u}^{\otimes2}}\:^\mathsf{T},\dots ,\:{\Bar{u}^{\otimes N_C}}\:^\mathsf{T}\right]^\mathsf{T},\label{eq:def_of_x}
    \end{align}
    where $N_C$ is the number of Carleman linearization steps. This vector is evolved by the linear equation
    \begin{align}
    \frac{dz}{dt}=Az+b,\quad \quad z(0)=\left[\Bar{u}_{\text{in}}^\mathsf{T},\:\Bar{u}_{\text{in}}^{\otimes2}\:^\mathsf{T},\dots ,\:\Bar{u}_{\text{in}}^{\otimes N_C}\:^\mathsf{T}\right]^\mathsf{T}.\label{eq:carlemanlinearizedeq}
    \end{align}
   For $N_C\to\infty$, Eq. \eqref{eq:carlemanlinearizedeq} encodes Eq. \eqref{eq:krovidef} exactly, meanwhile for finite $N_C$ it only approximates it. $A$ and $b$ can be constructed from $\Bar{F}^{(2)},\:F^{(1)}$ and $\Bar{F}^{(0)}$ as 
    \begin{align}
    \frac{d}{dt}\left[\begin{array}{c}
    z_1 \\
    z_2 \\
    z_3\\
    \vdots \\
    z_{N_C-1}\\[5pt]
    z_{N_C}
    \end{array}\right] = 
    \begin{pmatrix}
    A_1^1  &A_2^1 & & & & \\
    A_1^2  &A_2^2 &A_3^2&&&\\
    &A_2^3 &A_3^3 &A_4^3&&\\
     & &\ddots&\ddots&\ddots&\\
    & & & A_{N_C-2}^{N_C-1}&A_{N_C-1}^{N_C-1} &A_{N_C}^{N_C-1}\\[5pt]
    & & & &A_{N_C-1}^{N_C} &A_{N_C}^{N_C}
    \end{pmatrix}
    \left[\begin{array}{c}
    z_1 \\
    z_2 \\
    z_3\\
    \vdots \\
    z_{N_C-1}\\[5pt]
    z_{N_C}
    \end{array}\right]+
    \left[\begin{array}{c}
    \Bar{F}^{(0)} \\
    0 \\
    0\\
    \vdots \\
    0\\[5pt]
    0
    \end{array}\right],
    \label{eq:carlemanmatrix}
    \end{align}
    with $z_k=\Bar{u}^{\:\otimes k}$ as above. $A$ is a $d_A\equiv\left(d^{N_C+1}-d \right)/(d-1)$ dimensional, square and block-tridiagonal matrix. 
    The matrices inside are
       \begin{align}
    \begin{split}
    A_{j-1}^j&\equiv\Bar{F}^{(0)}\otimes\mathbb{I}^{\otimes j-1}+ \mathbb{I}\otimes \Bar{F}^{(0)} \otimes \mathbb{I}^{\otimes j-2}+\dots + \mathbb{I}^{\otimes j-1}\otimes \Bar{F}^{(0)},\\
    A_j^j&\equiv F^{(1)}\otimes\mathbb{I}^{\otimes j-1}+ \mathbb{I}\otimes F^{(1)} \otimes \mathbb{I}^{\otimes j-2}+\dots + \mathbb{I}^{\otimes j-1}\otimes F^{(1)} ,\\
    A_{j+1}^j&\equiv\Bar{F}^{(2)}\otimes\mathbb{I}^{\otimes j-1}+ \mathbb{I}\otimes \Bar{F}^{(2)} \otimes \mathbb{I}^{\otimes j-2}+\dots + \mathbb{I}^{\otimes j-1}\otimes \Bar{F}^{(2)}.
    \end{split}
    \end{align}

    \item The analytical solution of Eq. \eqref{eq:carlemanlinearizedeq} is
    \begin{equation}
  z(T)=\exp(AT)z(0)+\left(\int_0^T\exp(As)ds\right)b = \exp(AT)z(0) +T\sum_{j=0}^\infty \frac{(AT)^j}{(j+1)!}b.
    \label{eq:analytical_linear_sol}
    \end{equation}
    Time is discretized with increments of $h$ such that $t_l=lh$ with $l=0,\dots,m= T/h $, where for simplicity we assume that $h$ and $T$ are set so that the number of timesteps $m$ is an integer. The algorithm approximates Eq. \eqref{eq:analytical_linear_sol} at every timestep as
    \begin{align}
    y_{l+1}=T_k(Ah)y_l +S_k(Ah)hb,\label{eq:stepping}
    \end{align}
    where $y_l$ approximates $z(t_l)$. The matrices in the above equation are Taylor series truncated at level $k$ as
    \begin{align}
    T_k(w) \equiv \sum_{j=0}^k\frac{w^j}{j!}\quad ,\quad S_k(w)\equiv\sum_{j=1}^k\frac{w^{j-1}}{j!}.\label{eq:T_k}
    \end{align}
    Then Eq. \eqref{eq:stepping} is implemented as a linear system of equations 
    \begin{align}
    L|y\rangle=|\psi_{\text{in}}\rangle,\label{eq:linearsystemofeqs}
    \end{align}
    with a matrix $L$ formulated as
    \begin{align}
    L&=\mathbb{I}-N\label{eq:matrix_L}\\
    N&=\sum_{i=0}^m|i+1\rangle \langle i|\otimes M_2(\mathbb{I}-M_1)^{-1}+\sum_{i=m+1}^{m+p-1}|i+1\rangle \langle i|\otimes \mathbb{I}\nonumber\\
    M_1 &= \sum_{j=0}^{k-1}|j+1\rangle\langle j |\otimes \frac{Ah}{j+1} \nonumber\\
    M_2 &= \sum_{j=0}^k |0\rangle \langle j| \otimes \mathbb{I}.\nonumber
    \end{align}
    The RHS of Eq. \eqref{eq:linearsystemofeqs} is
    \begin{align}
    |\psi_{\text{in}}\rangle &=\frac{1}{\sqrt{\norm{z_0}^2+mh^2\norm{b}^2}}\left(|0,0,z_0\rangle +h\sum_{i=0}^{m-1}|i,1,b\rangle \right).\label{eq:psi_in}
    \end{align}
    The first register, indexed by $i$, encodes the index of the discrete time step in the computational basis.  The second register, indexed by $j$, represents the index of the truncated Taylor series term (corresponding to the sub-time-steps of the integrator), while the third register stores the state vectors on which the operator $A$ acts. In the above state $|z_0\rangle$ encodes the initial condition of the linearized system~\cite{liu}
    \begin{align}
    |z_0\rangle = \frac{1}{\sqrt{V}}\sum_{j=1}^{N_C} \norm{\Bar{u}_{\text{in}}}^j |j\rangle\otimes |\Bar{u}_{\text{in}}\rangle^{\otimes j} \otimes |0\rangle^{\otimes N_C-j},\quad \text{where}\quad V=\sum_{j=1}^{N_C} \norm{\Bar{u}_{\text{in}}}^{2j}. \label{eq:zin}
    \end{align}
    The $|0\rangle$ state in the third term of the tensor product above is $d$ dimensional. This ensures that the dimension of $|\Bar{u}_{\text{in}}\rangle^{\otimes j} \otimes |0\rangle^{\otimes N_C-j}$ is fixed to be $d^{N_C}$ $\forall\: j$ for mathematical consistency. The state preparation of $|\psi_{\text{in}}\rangle$ is described in Ref. \cite{liu}.

    \item The last step is inverting Eq. \eqref{eq:linearsystemofeqs} to obtain the state
    \begin{align}
    \begin{split}
    |y\rangle&=L^{-1}|\psi_{\text{in}}\rangle\\
    & = \sum_{i=0}^{m}|i,0,y_i\rangle +\sum_{i=m+1}^{m+p-1}|i,0,y_m\rangle.\label{eq:result_state}
    \end{split}
    \end{align}
    Here the point of having the parameter $p$ becomes apparent: it increases the amplitude of the $|y_m\rangle$ states we are looking for. In fact, the first $d$ components of $|y_m\rangle$, which we call $|y_{1,m}\rangle$, is level 1 in the Carleman embedding, hence it encodes $\bar{u}(T)$, the rescaled solution of \eqref{eq:krovidef}. To ensure that the success probability for measuring $|y_{1,m}\rangle$ is constant, further amplitude amplification is performed on Eq. \eqref{eq:result_state}.
    
\end{enumerate}

\subsection{Oracle access model}\label{sec:oracles}
We now specify the oracle access model assumed by the quantum algorithm, in the way that is conventional in prior work on differential equations \cite{liu,berry1,berry2,partial,childs}. The input consists of the rescaled matrices $\bar F^{(2)}$, $F^{(1)}$, the rescaled inhomogeneous vector $\bar F^{(0)}$, and the rescaled initial condition $\bar u_{\mathrm{in}}$. For each matrix $M \in \{\bar F^{(2)},F^{(1)}\}$, we assume standard sparse-access oracles,
\begin{align}
\begin{split}
O_{M,\mathrm{loc}}|r,l\rangle &= |r,c_M(r,l)\rangle,\\
O_{M,\mathrm{val}}|r,c\rangle|0\rangle &= |r,c\rangle|M_{rc}\rangle,
\end{split}
\end{align}
where $c_M(r,l)$ is the column index of the $l$-th nonzero entry in row $r$, for $l=1,\ldots,s_M$. Furthermore, $M_{rc}$ is an approximate binary representation of the $(r,c)$ matrix entry of $M$. Similarly for the inhomogeneous term, we assume state-preparation access to the normalized vector
\begin{align}
O_{\bar F^{(0)}}|0\rangle
=
\frac{|\bar F^{(0)}\rangle}{\|\bar F^{(0)}\|}.
\end{align}
Together with the state-preparation oracle
\begin{align}
O_{\mathrm{in}}|0\rangle
=
\frac{|\bar u_{\mathrm{in}}\rangle}{\|\bar u_{\mathrm{in}}\|},
\end{align}
this provides the primitive input data from which the Carleman-linearized system is constructed. In particular, the matrix $A$, the vectors $b$ and $z(0)$, and hence the linear system in Eq.~\eqref{eq:linearsystemofeqs} are obtained by standard matrix arithmetic on the blocks built from $\bar F^{(2)}$, $F^{(1)}$, $\bar F^{(0)}$, and $\bar u_{\mathrm{in}}$. In the complexity analysis below, we therefore treat the construction of the required block encodings and the resulting matrix operations as consequences of these oracle-access assumptions.

\subsection{Errors}\label{errors_section}
There are three sources of errors in the quantum algorithm in Section~\ref{subsec:qalgo}. 
Firstly, the truncation of the Carleman linearization at level $N_C$ in Eqs. (\ref{eq:def_of_x}-\ref{eq:carlemanlinearizedeq}) introduces an error $\delta$ relative to the exact solution, defined as 
\begin{align}
\norm{z_1(T)-\Bar{u}(T)}\leq \delta \norm{\Bar{u}(T)},\label{eq:carleman_error_bound}
\end{align}
provided that $N_C$ is chosen according to \cite[Lemma 17]{krovi} as
\begin{align}
N_C \geq \left\lceil \frac{2\log(T\norm{\Bar{F}^{(2)}}/\delta\norm{\Bar{u}(T)})}{\log(1/\norm{\Bar{u}_{\text{in}}})}\right\rceil,\label{eq:parameters_chosen_geq}
\end{align}
for a desired $\delta$. Secondly, the truncation of the temporal Taylor series in Eq. \eqref{eq:stepping} at level $k$ introduces a relative error $\delta'$ on the Carleman linearized state vector as
\begin{align}
\norm{y_m - z(T)}\leq \delta'\norm{z(T)},\label{eq:Taylor_error_2}
\end{align}
where $z(T)$ is the exact solution of the linear system from Eq. \eqref{eq:analytical_linear_sol} and $y_m$ approximates it. The above inequality is conditioned on choosing $k$ in terms of $\delta'$ according to \cite[Theorem 3]{krovi} as
\begin{align}
(k+1)!\geq \frac{me^3}{\delta'}\left(1+\frac{Te^2\norm{b}}{\norm{z(T)}}\right),\label{eq:Taylor_error_1}
\end{align}
and on having $\norm{Ah}\leq 1$ (which is satisfied due to the choice of $h$ later). Note that Eq. \eqref{eq:Taylor_error_2} is also a bound on the temporal error of the first block components of $y_m $, namely $y_{1,m}$. Hence, by bounding $\norm{z(T)}$ on the RHS of Eq. \eqref{eq:Taylor_error_2} as $\norm{z(T)}\leq (1+\delta)\sqrt{N_C}\norm{\bar{u}(T)}$ via \cite[Theorem 8]{krovi}, we may rewrite it as
\begin{align}
\norm{y_{1,m} - z_1(T)}\leq \delta'(1+\delta)\sqrt{N_C}\norm{\bar{u}(T)}.\label{eq:rewritten_taylor_error}
\end{align}
Combining Eq. \eqref{eq:carleman_error_bound} with Eq. \eqref{eq:rewritten_taylor_error} we obtain the bound 
\begin{align}
\begin{split}
\norm{y_{1,m}-\bar{u}(T)} &= \norm{y_{1,m}-z_1(T) +z_1(T) - \bar{u}(T)}\\
&\leq \norm{y_{1,m}-z_1(T)} + \norm{z_1(T)-\bar{u}(T)}\\
&\leq \Big(  \delta'(1+\delta)\sqrt{N_C}+\delta \Big)\norm{\bar{u}(T)}.
\end{split}
\end{align}
Finally, the QLSA introduces a purely quantum error denoted by $\varepsilon_q$. We therefore choose $\delta,\,\delta^\prime$ such that $\delta'(1+\delta)\sqrt{N_C}+\delta\leq \varepsilon_q/2$, since then according to \cite[Theorem 8]{krovi}, the full error on the normalized state also satisfies
\begin{align}
\begin{split}
\norm{\frac{y_{1,m}}{\norm{y_{1,m}}} - \frac{\Bar{u}(T)}{\norm{\Bar{u}(T)}}}\leq \varepsilon_q.
\label{eq:qstate}
\end{split}
\end{align}

\subsection{Complexity of the QLSA}\label{QLSAcomplexity}
The Carleman linearized system is put into the form of a system of linear equations after the time discretization, as given in Eq. \eqref{eq:linearsystemofeqs}. The QLSA applied to that linear system produces the state $L^{-1}|\psi_{\text{in}}\rangle$ with an error of $\varepsilon_q$ and has a query complexity of
\begin{align}
\mathcal{O}\left(s_Ak\kappa(L)
\text{polylog}\left[k,m,d_A,\kappa(L),\frac{1}{\varepsilon_q}
\right]
\right),
\label{eq:qlsa_querycomplexity}
\end{align}
and a gate complexity that is larger by a factor of
\begin{align}
\mathcal{O}\left(
\text{polylog}\left[k,m,\frac{1}{\varepsilon_q}
\right]
\right),
\label{eq:qlsa_gatecomplexity_factor}
\end{align}
as it was shown in \cite[Theorem 6]{krovi}. In the above, $s_A$ is the sparsity of $A$, $d_A$ is its dimension, $\kappa(L)$ is the condition number of the linear system, $k$ is the truncation level of the temporal Taylor series and $m$ is the number of time-steps. We now apply the QLSA to the Carleman linearized system.

\subsection{Complexity of the nonlinear solver}\label{gatecomplexity}
The quantum algorithm first prepares $|\psi_{\text{in}}\rangle$ from Eq. \eqref{eq:psi_in}, then applies the QLSA from above to produce $L^{-1}|\psi_{\text{in}}\rangle$ and finally amplifies the amplitude of the $|y_{1,m}\rangle$ states. If $\mu(F^{(1)})<0$ and $R<1$, the quantum algorithm produces a quantum state satisfying Eq. \eqref{eq:qstate} with a query complexity \cite[Theorem 8]{krovi}
\begin{align}
\mathcal{O}\left(\sqrt{N_C(\delta)}g_u sN_C(\delta) T\norm{A}\frac{\log \Omega(\delta',\delta)}{\log  \log \Omega(\delta',\delta)}
\text{polylog}\left[\log \Omega(\delta',\delta),T\norm{A},d^{N_C(\delta)},\frac{1}{\varepsilon_q}
\right]
\right),
\label{eq:querycomplexity}
\end{align}
and a gate complexity that is larger by a factor of 
\begin{align}
\mathcal{O}\left(
\text{polylog}\left[\log \Omega(\delta',\delta),T\norm{A},\frac{1}{\varepsilon_q}
\right]
\right).
\label{eq:gatecomplexity_factor}
\end{align}
\begin{enumerate}
\item The above complexities assume that the following internal parameters are picked: 
\begin{align}
\begin{split}
h&=\frac{T}{\lceil T\norm{A}\rceil},\quad m=p=\frac{T}{h} = \lceil T\norm{A}\rceil,\quad \delta+(1+\delta)\delta'\sqrt{N_C(\delta)}\leq \varepsilon_q/2,\\
k(\delta',\delta)&=\left\lceil \frac{2\log\Omega(\delta',\delta)}{\log \log \Omega(\delta',\delta)}\right \rceil,\quad \Omega(\delta',\delta)=\frac{e^3T\norm{A}}{\delta'}\left(1+\frac{Te^2\norm{\bar{F}^{(0)}}}{\norm{\bar{u}(T)}}\right),\\
N_C(\delta) &= \left\lceil \frac{2\log(T\norm{\Bar{F}^{(2)}}/\delta\norm{\Bar{u}(T)})}{\log(1/\norm{\Bar{u}_{\text{in}}})}\right\rceil.
\label{eq:parameters_chosen}
\end{split}
\end{align}
    \item $\Omega(\delta',\delta)$ above is defined such that $(k(\delta',\delta)+1)!>\Omega(\delta',\delta)$ and therefore Eq. \eqref{eq:Taylor_error_1} is satisfied. We included the explicit $\delta'$ and $\delta$ dependence for clarity, noting that the $\delta$ dependence originates from the dependence of the norms in $\Omega(\delta',\delta)$ on $N_C(\delta)$.
    \item  Similarly, setting $N_C(\delta)$ according to Eq. \eqref{eq:parameters_chosen} ensures that Eq. \eqref{eq:carleman_error_bound} is satisfied.
    \item  The choice for $\delta$ and $\delta'$ in Eq. \eqref{eq:parameters_chosen} makes the $l_2$ error of the solution state $\varepsilon_q$ as shown in Eq. \eqref{eq:qstate}.
    \item The denominator in $k(\delta',\delta)$, namely $\log \log \Omega(\delta',\delta)$, vanishes asymptotically when $k(\delta',\delta)$ is in logarithms in Eqs. (\ref{eq:qlsa_querycomplexity}-\ref{eq:qlsa_gatecomplexity_factor}). Hence the polylog$(\cdot)$'s contain no $\log \log \Omega(\delta',\delta)$ terms. 
    \item  The factor $sN_C(\delta)$ in Eq. \eqref{eq:querycomplexity} originates from the sparsity of $A$ being $s_A=3sN_C(\delta)$, where $s$ denotes the sparsity of $\Bar{F}^{(2)},F^{(1)},\Bar{F}^{(0)}$.
    \item $\norm{A}$ can be bounded as
\begin{align}
\norm{A}\leq N_C(\delta)\left( \norm{\Bar{F}^{(0)}} +\norm{F^{(1)}}+\norm{\Bar{F}^{(2)}}\right).\label{eq:Anorm}
\end{align}
\item The condition number of $L$ is asymptotically bounded as $\kappa(L)=\mathcal{O}(m+p)=\mathcal{O}(T\norm{A})$ using \cite[Theorem 4]{krovi}.
\item The dimension of $A$ is $d_A=\mathcal{O}(d^{N_C(\delta)})$. The contribution of this term to the query complexity is hence polynomial in $N_C(\delta)$.
\item 
The factor $\sqrt{N_C(\delta)}g_u$ in Eq. \eqref{eq:querycomplexity} is the result of applying amplitude amplification on the $|y_{1,m}\rangle$ states, as the probability to measure the $y_{1,m}$ states can be lower bounded as
\begin{align}
P[\text{measure } y_{1,m}]\geq \mathcal{O}\left(\frac{1}{N_Cg_u^2}\right),\quad\quad g_u\equiv \frac{\norm{u_{\text{in}}}}{{\norm{u(T)}}}.
\end{align}
\item The the polylog$(\cdot)$ expressions are polynomials of the logarithms of their inputs, and this polynomial can only be determined entirely during the actual implementation of the quantum algorithm. 
\item  Note that $s$ appears in Ref. \cite{krovi} within the polynomial term, but it is actually linear only. That is because the underlying QLSA used is linear in $s$, as indicated in Eq. \eqref{eq:qlsa_querycomplexity}.
\end{enumerate}

\section{Mapping onto the quantum algorithm}\label{sec:mapping}

We now introduce the mapping between our discretized Vlasov equation, as formulated in Eqs. (\ref{eq:1species_discretized_a}-\ref{eq:1species_discretized_d}), onto the ODE system in Eq. \eqref{eq:krovidef}.
The contents of the $\mathcal{O}(f^2)$ line encode the nonlinearity, hence they will determine $F^{(2)}$. Similarly, the  $\mathcal{O}(f^1)$ lines determine $F^{(1)}$ and the $\mathcal{O}(f^0)$ one determines $F^{(0)}$.

The key idea to carry out this mapping is to place the dynamical variables in a vector and treat that as the state vector $u$. In the Gauss's law case it is the vectorization of the distribution matrix $f$ from Eq. \eqref{eq:distributionmatrix}, while in the Ampere's law case the $E_i$ field variables would also be part of $u$. The latter is discussed in Section \ref{sec:ampere}. This sort of `packaging' can be performed in many different ways, but here we employ row-major vectorization, which operation is explained in Appendix \ref{vectorization}.
When coupling to Gauss's law, this produces the state vector $u$ as
\begin{align}
u= \text{vec}(f)
=
\left[ f_{1,1}, \:  \cdots, \: f_{1,N_v}, \: f_{2,1}, \: \cdots, \: f_{2,N_v},\: \dots, \: f_{N_x,N_v} \right]^\mathsf{T},
\label{eq:uvecdef1species}
\end{align}
with the length $N=N_xN_v$ being the total number of grid-points. We show in Section \ref{section:discussion} that the choice of the vectorization method does not affect the complexity of the problem.

After the construction of $u$, the entries of the matrices $F^{(2)},\:F^{(1)},\:F^{(0)}$ can be obtained procedurally from the finite difference equations as explained next.

\subsection{Method to find the $F^{(i)}$ matrices parameterizing the nonlinear ODE}\label{sec:method_to}
In Appendix \ref{appendix:vectorization_of_u_gauss} we state the so-called conversion formulas that map between entries of $f$ and $u$, as well as between quadratic monomials of those entries. Upon substituting the conversion formulas (\ref{eq:fijfromun},\ref{eq:u2fromff}) into the finite difference equations (\ref{eq:1species_discretized_a}-\ref{eq:1species_discretized_d}) and simplifying the expressions, one obtains an evolution equation of the form
\begin{align}
\frac{du_n}{dt}=\sum_{m\in S^{(n)(2)}}C_{m}^{(n)(2)}u^{\otimes2}_m+\sum_{m\in S^{(n)(1)}}C_{m}^{(n)(1)}u_m+\sum_{m\in S^{(n)(0)}}C_{m}^{(n)(0)},\label{eq:linearcombination}
\end{align}
with coefficients $C_{m}^{(n)(2)},\:C_{m}^{(n)(1)},\:C_{m}^{(n)(0)}$. The first two encode a \emph{weighting} for the elements of $u^{\otimes2}$ and $u$, respectively, while $C_{m}^{(n)(0)}$ encodes the inhomogeneity. The sets $S^{(n)(2)}$ and $S^{(n)(1)}$ describe \emph{which elements} of the mentioned vectors take part in evolving $u_n$. The superscript $(n)$ denotes which element of $u$ is being evolved.

The key point is that the above sums are \emph{linear combinations} of the elements of $u^{\otimes2}$ and $u$. These linear combinations can be identified as matrix multiplications with suitable matrices $F^{(2)}$ and $F^{(1)}$. By also denoting the inhomogeneous terms with a matrix (a vector to be exact), called $F^{(0)}$, one obtains an ODE system of the form of Eq. \eqref{eq:krovidef} where the matrices have entries
\begin{subequations}
\begin{align}
\left[F^{(2)}\right]_{n,k}&=\sum_{m\in S^{(n)(2)}}C_{m}^{(n)(2)}\delta_{km},\\
\left[F^{(1)}\right]_{n,k}&=\sum_{m\in S^{(n)(1)}}C_{m}^{(n)(1)}\delta_{km},
\end{align}
\end{subequations}
and the vector
\begin{equation}
\left[F^{(0)}\right]_{n}=\sum_{m\in S^{(n)(0)}}C_{m}^{(n)(0)}.
\end{equation}
The takeaway is that the above analytical maps are directly deducible from the finite difference equations in a procedural way, once the conversion formulas are known.

\subsection{The matrix entry maps}\label{section:matrices}
The above procedure results in maps for the entries of the $F^{(k)}$ matrices. In Appendix \ref{appendix:entries} all three are written out explicitly with the Gauss's law coupling, and $F^{(1)}$ with the Ampere's law coupling.

\begin{figure}[h!]
    \centering
\includegraphics[width=5.65 in]{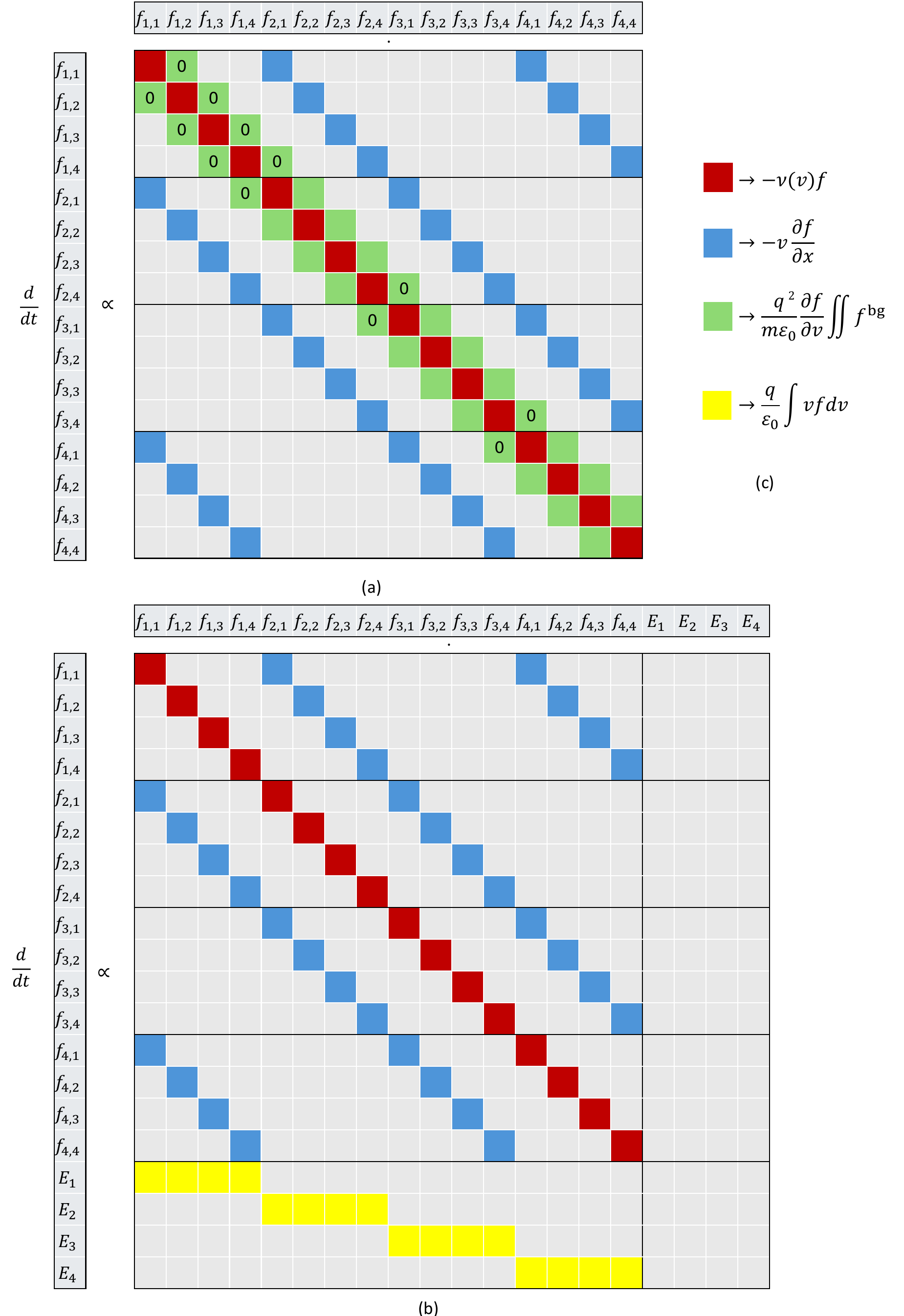}
\caption{Illustration of the $F^{(1)}$ matrix for a $(4,4)$ grid. (a): with Gauss's law; (b): with Ampere's law; (c): color codes for terms from Eqs.~(\ref{eq:onespecisesonedimension}-\ref{eq:ampereevolution}). On the left of (a) and (b), the evolved vector $u$ is written out explicitly, meanwhile on top, $u$ represents how the matrix multiplication mixes its elements. All gray elements are zero. Cells with the same colors have generally different numerical values.}
\label{fig:double_F1}
\end{figure}

The linear contribution is partitioned into two parts as $F^{(1)}\equiv F^{(1a)}+F^{(1b)}$. 
Let $F^{(1a)}$ encode the collisional part of the linear evolution, Eq. \eqref{eq:1species_discretized_c}, and $F^{(1b)}$ encode the rest of the linear contribution, Eq. \eqref{eq:1species_discretized_b}. An identical mapping for Ampere's law is discussed in Section \ref{sec:ampere}. The full $F^{(1)}$ matrices are visually represented on Fig. \ref{fig:double_F1} for a (4,4) grid. Both couplings result in a diagonal $F^{(1a)}$ and off-diagonal $F^{(1b)}$, however the inner structures of the matrices are different. 
In  Fig. \ref{fig:double_F1}~(a), which is the Gauss's law version, $F^{(1b)}$ is antisymmetric, otherwise known as skew-symmetric. This can be analytically confirmed from the maps given in Eqs. (\ref{eq:F1bar}-\ref{eq:F1hatmap}). 
In  Fig. \ref{fig:double_F1}~(b), which is the Ampere's law version, the $F^{(1b)}$ contribution has no symmetry structure in it. This version has empty columns, which carry special importance form the point of view of the Carleman linearization as it is investigated later in Section \ref{sec:ampere}.

\begin{figure}[h!]
    \centering
\includegraphics[width=5.65 in]{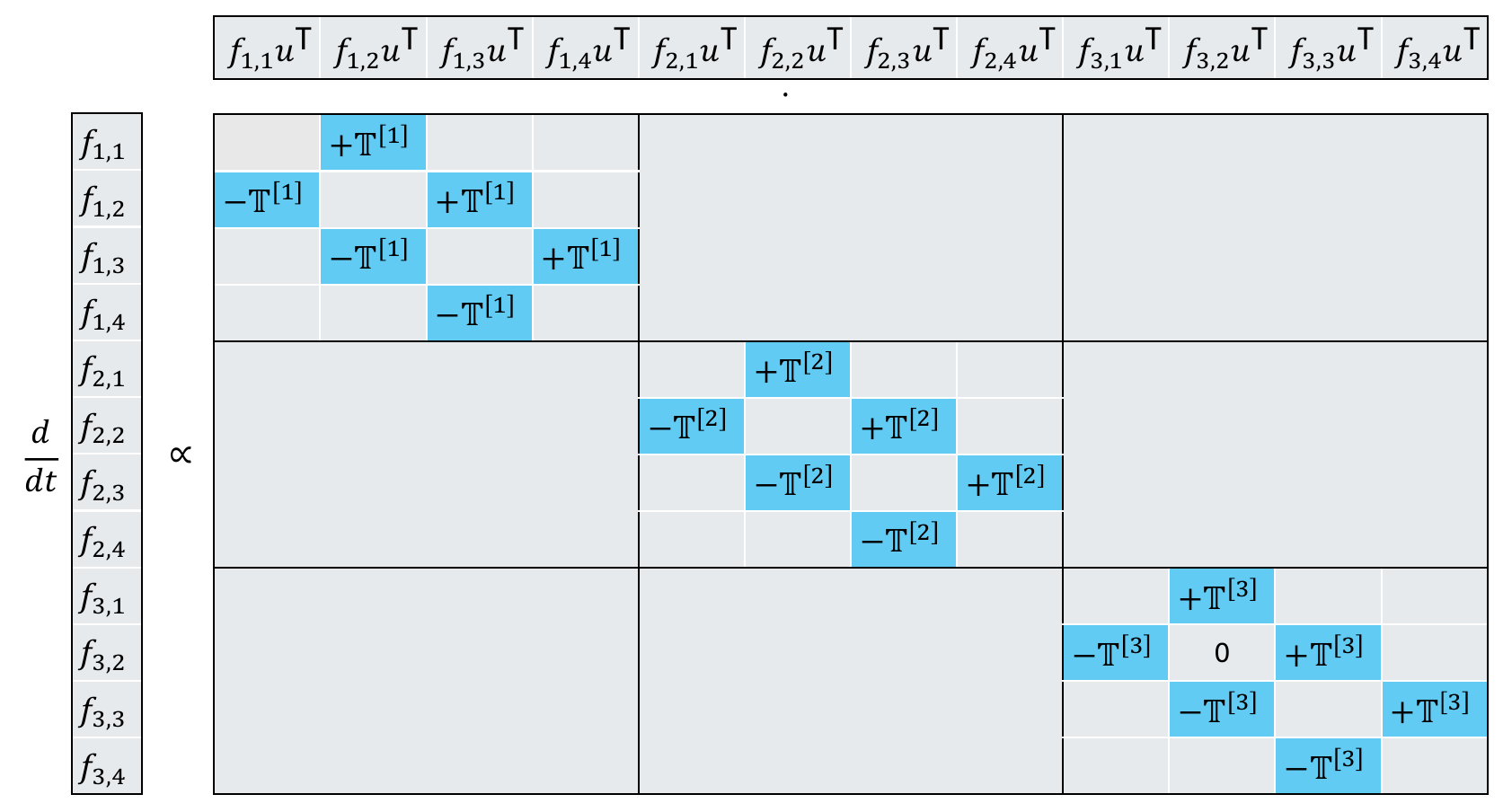}
\caption{Illustration of the Gauss's law version of $F^{(2)}$. The middle matrix is $F^{(2)}$ up to a constant factor of $-q^2\Delta x/(8m_e\varepsilon_0)$, for a $(3,4)$ grid. The $\mathbb{T}^{[i]}$ vectors encode the double integral stencil as are defined in Appendix~\ref{F2subsection}. On the left, the evolved vector $u$ is written out explicitly  The small cells are row vectors with length $N=12$. On top $u^{\otimes2}$ represents how the matrix multiplication mixes its elements. Gray elements are 0.}
\label{fig:F2_Gauss_fig}
\end{figure}

We derive $F^{(2)}$ in Appendix~\ref{F2subsection} with Gauss's law as the coupling. Fig. \ref{fig:F2_Gauss_fig} provides an illustration for a (3,4) grid. We are now ready to analyze the convergence criteria of the quantum algorithm given these matrices. 

\section{Investigating convergence}\label{convergence}
\subsection{Strategy}
The convergence of the quantum algorithm from Section \ref{section:quantum_algo} is investigated in the limit of $N_x,N_v\gg1$. For the sake of the convergence analysis, since $\nu_0$ (from Eq. \eqref{eq:defofnu}) controls the strength of the dissipation in the system, it is beneficial to treat it similarly to $N_x\:,N_v$, which are \emph{control parameters}. This means that the scaling of each mathematical expression is realized with respect to these three parameters. On the other hand variables such as $x_{\text{max}},\:v_{\text{max}},\:\mathcal{N}$ and $b$ are treated as fixed. Moreover, the function $h(v)$ (from Eq. \eqref{eq:defofnu}) is also treated as fixed and independent of the control parameters.

Finally, $N_v$ is assumed to be even from now on. These assumptions simplify the computation of certain terms to some extent as it is shown later. Remember that Eq. \eqref{eq:xcoordinate} fixes $\Delta x$ from $N_x$ and $x_{\text{max}}$ as well as that Eq. \eqref{eq:vcoordinate} fixes $\Delta v$ from $N_v$ and $v_{\text{max}}$.

We shall show later that the value of $\nu_0$ puts a restriction on the maximum value $N_v$ may take. Hence, within the calculation below we also treat $\nu_0$ as a potentially large value.

\subsection{Ingredients for the $R$ parameter}\label{sec:ingredients_subsec}
In Appendix \ref{appendix:F2_norm} we compute the exact $\Vert F^{(2)}\Vert$ and in Appendix \ref{section:F0_norm} we compute the asymptotic scaling of $\Vert F^{(0)}\Vert$.

As for $\mu(F^{(1)})$, recall from Subsection \ref{section:matrices} the decomposition $F^{(1)}=F^{(1a)}+F^{(1b)}$, where $F^{(1a)}$ is diagonal and $F^{(1b)}$ is antisymmetric. Then we have
\begin{align}
\begin{split}
\mu\left(F^{(1)}\right)&=\lambda_{\max}\left\{\frac{F^{(1a)}+F^{(1b)}+\left(F^{(1a)}+F^{(1b)}\right)^T}{2}\right\}\\
&=\lambda_{\max}\left\{F^{(1a)}\right\}\\
&=\max_{j=1,\dots,N_v}\left\{-\nu(v_j)\right\}\\
&\leq -\nu_0.\label{eq:mubound}
\end{split}
\end{align}
Therefore $\mu(F^{(1)})\leq -\nu_0<0$, as required. We see that without collisions, this condition would not be satisfied in our model. The simplicity of the above calculation is the advantage behind using the Krook collision operator.

The choice of possible initial conditions is restricted by two factors. Firstly, $\norm{u_{\text{in}}}$ needs to be exactly computable as it appears multiple times, in both numerators and denominators of expressions. Secondly, $u_{\text{in}}$ must be more \emph{concentrated}\footnote{Concentration in phase space refers to $f(x,v)$ being large in the vicinity of certain $x$ and $v$ values, and close to $0$ elsewhere.} in phase space than the Maxwellian (Gaussian) thermal state $f^{M}(v)$ encoded in $F^{(0)}$. This is fundamentally because the ratio $\Vert F^{(0)}\Vert /\norm{u_{\text{in}}}$ must decrease as the system size $(N_x,N_v)$ grows. 
Recall from Eq. \eqref{eq:1species_discretized_d} and correspondingly Eq. \eqref{eq:F0map} that the Gaussian in $F^{(0)}$ is weighted by the values of $\nu(v)$. This weighting can be thought of as being `cancelled' by the $\nu_0$ in $\mu(F^{(1)})$ in the denominator of $R$ as shown in the above paragraph. With this cancellation the ratio $\Vert F^{(0)}\Vert /\norm{u_{\text{in}}}$ becomes the norm of a Gaussian over the norm of $u_{\text{in}}$. The more evenly distributed the elements of a vector are, the smaller its $l_2$ norm becomes. Consequently, $u_{\text{in}}$ must be more concentrated in phase space than $f^{M}(v)$.

The above two conditions are satisfied for a number of initial conditions. We choose a scenario in which two spatially uniform monoenergetic beams with opposite, definite velocities collide head-on. This occurs in nature in the form of astrophysical jet collisions \cite{ferrari_plasma_jet,physics_of_space_plasmas}. Then we have $f_{ij}(t=0)\propto \delta_{j,J}+\delta_{j,N_v-J+1}$, which encodes the two beams with velocities $v_J$ and $-v_J$ for some $J$. The lack of $i$ dependence ensures uniformity in $x$. This may be illustrated as
\begin{align}
u_{\text{in}}\propto \text{vec}
\left({\begin{array}{ccccccccccc}
0 & \cdots & 0 & 1 & 0 & \cdots & 0 & 1 & 0 & \cdots & 0 \\
\vdots & \ddots & \vdots & \vdots & \vdots & \ddots & \vdots & \vdots & \vdots & \ddots & \vdots \\
0 & \cdots & 0 & 1 & 0 & \cdots & 0 & 1 & 0 & \cdots & 0 \\
\end{array}}\right),\label{eq:f(0)_matrix}
\end{align}
where the $1$'s are in the $J$'th and $(N_v-J+1)$'th columns. The proportionality constant can be fixed from requiring that $\iint f_{ij}(t=0)=\mathcal{N}$ on the full grid, using the trapezoidal rule for integration. This results in the map given in Appendix \ref{section:u0_norm}.

Upon substituting $\mu(F^{(1)})$ from Eq. \eqref{eq:mubound} and $\Vert F^{(2)}\Vert,\: \Vert F^{(0)}\Vert,\: \norm{u_{\text{in}}}$ from Appendix \ref{appendix:norm_derivations} into the $R$ parameter given in Eq. \eqref{eq:R_def}, in the large $N_x,\:N_v$ limit one obtains the full asymptotic expression
\begin{align}
\lim_{N_x,\:N_v \gg 1} R =\frac{q^2 \mathcal{N}}{2\sqrt{2}m_e\varepsilon_0 v_{\max}} \frac{N_v^{3/2}}{\nu_0}= \mathcal{O}\left(\frac{N_v^{3/2}}{\nu_0}\right). \label{eq:nu_0_scaling}
\end{align}
As a result of our choice for $u_{\text{in}}$, the factor $ \Vert F^{(0)}\Vert / [\norm{u_{\text{in}}} \mu(F^{(1)})] $  entirely vanishes in the asymptotic limit. 
If that was not the case, then regardless of the value of $\nu_0$, the $R$ parameter would have either a constant $\mathcal{O}(1)$ contribution from the mentioned factor, or would explicitly increase with system size $(N_x,N_v)$.

\subsection{Connection to plasma physics}\label{sec:connection_to_plasma}
To gain insight into the level of restriction on plasma parameters Eq. \eqref{eq:nu_0_scaling} requires, we will employ a simple model for the collision frequency that is commonly used in plasma physics \cite{plasma2}. It reads
\begin{align}
\nu_0 = \frac{q^4\Bar{n}\log(\Lambda)}{(4\pi\varepsilon_0)^2 m_e^{1/2}(\mathcal{T}k_B)^{3/2}},\label{eq:actualnu0}
\end{align}
where $\log(\Lambda)$ is the Coulomb logarithm, whose value is around $\log(\Lambda)\approx 10$. We can combine this with Eq. \eqref{eq:nu_0_scaling}, set $R<1$ and $\mathcal{N}=\bar{n}x_{\max}$. Note that we choose $v_{\text{max}}=10\cdot v_p$, with $v_p=1/\sqrt{b}$ being the most probable velocity and $b=m_e/(2k_B\mathcal{T})$ being the decay factor of the Maxwellian. Then this gives us the bound
\begin{align}
N_v  \lesssim \left(
\frac{25}{\pi^2}\frac{q^2}{\varepsilon_0 k_B}\frac{1}{x_{\max}\mathcal{T}}
\right)^{2/3}.\label{eq:N_v_restriction}
\end{align}

We will first use typical physical values of astrophysical plasma within interstellar medium. For this setup our BCs and the $(1+1)$ dimensional approximation is relatively accurate when considering the flow of ionized matter far from its sources, e.g. between distant stars. The concrete example we shall use is warm ionized medium (WIM) made up of electrons and ionized hydrogen, which takes up 20-50 \% of the interstellar matter of the Milky Way. Its typical temperature is $\mathcal{T}=8000$ K \cite{plasmastuff}. Additionally, we choose $x_{\text{max}}=1000$ km, which, using Eq. \eqref{eq:N_v_restriction}, gives
\begin{align}
N_v\lesssim 1.6\cdot 10^{-9},
\end{align}
which is many orders of magnitude below grid sizes of interest.

Another setup would be inertial confinement fusion with physical parameters in the order of $\mathcal{T}=50$ million K and $x_{\max}=10^{-4}$ m \cite{plasma2}, giving us
\begin{align}
N_v\lesssim 2.24\cdot 10^{-5},
\end{align}
which is again unrealistic. 

We may also try substituting a reasonable velocity grid size of $N_v\geq 100$ into Eq. \eqref{eq:N_v_restriction}. Using the numerical values of $q,\:\varepsilon_0$ and $k_B$, then gives the bound
\begin{align}
x_{\max}\mathcal{T}\lesssim 5.31\cdot 10^{-7}\:\text{mK}.
\end{align}
The typical value of the product of $x_{\max}$ and $\mathcal{T}$ in real-world setups are orders of magnitude above what this inequality would require.

Consequently, we conclude that the region of convergence of the quantum algorithm does not include scenarios of physical interest. In order to solve the system with reasonable grid sizes, collisions have to be unphysically strong in our model.

Finally, let us briefly discuss potential values of the final simulation time $T$ that are of scientific relevance. $T$ should be chosen to match the physical process of interest. For a transport or wave-propagation problem, such as the warm ionized medium example, a natural choice is the time required for information to cross the simulation domain, for example $T\sim x_{\max}/v_p$. For a relaxation problem dominated by collisions, one may instead choose $T$ to be a fixed small multiple of the collision time, for example $T=c/\nu_0$ with $c>1$ chosen according to the amount of relaxation one wishes to resolve. Thus, a quantitative estimate of $T$ depends on the specific plasma scenario and observable being targeted.

\section{Summary of algorithmic errors}\label{section:errors}
In this section we analyze the errors that arise when solving the finite difference equations, Eqs. (\ref{eq:1species_discretized_a}-\ref{eq:1species_discretized_d}), with the quantum algorithm discussed. We assume the plasma parameters are set such that the convergence criteria investigated in the above section are satisfied. The errors that arise will be compared to those of the classical solution derived in Appendix \ref{appendix:classical_error_complexity}. A total of four different errors are contained within the output state of the quantum algorithm, as shown in Table \ref{tab:errors}. Here, $\varepsilon_c$ denotes the phase-space discretization error associated with the finite grid, whereas $\varepsilon_q$ denotes the target quantum algorithmic error, that is set to include Carleman truncation, Taylor truncation, and QLSA error. The total error is therefore $\mathcal{O}(\varepsilon_c+\varepsilon_q)$.

\begin{table}[h]
\centering
\begin{tabular}{|c|cc|}
\hline
  \phantom{\Big|} \textbf{Source of error}                            & \multicolumn{2}{c|}{\textbf{Size}}                \\ \hline\hline 
\phantom{\Big|} Phase space discretization                                       & \multicolumn{2}{c|}{$\varepsilon_c\sim T/(N_x^2+N_v^2)$}         \\ \hline
\phantom{\Big|} Truncating Carleman linearization &  \multicolumn{1}{c|}{$\quad \delta\quad$}  &    \multirow{2}{*}{$\Rightarrow \delta+(1+\delta)\delta'\sqrt{N_C(\delta)}$}  \\ \cline{1-2}
 \phantom{\Big|}Truncating temporal Taylor series   &       \multicolumn{1}{c|}{$\quad\delta'\quad$} &            \\ \hline
\phantom{\Big|} QLSA                                       & \multicolumn{2}{c|}{$\varepsilon_q$}         \\ \hline
\end{tabular}
\caption{Summary of the errors that enter the final quantum state produced by the quantum algorithm applied to the discretized Vlasov equation.}
\label{tab:errors}
\end{table}

The finite difference derivatives and integral carry an (absolute) error of $\mathcal{O}(\Delta x^2+\Delta v^2 )$, which enters the $F^{(j)}$ matrices on the RHS of the quadratic ODE in Eq. \eqref{eq:krovidef}. It is furthermore directly pushed into the Carleman linearized evolution matrix $A$ in Eq. \eqref{eq:carlemanmatrix}. Then the system of linear equations described by the matrix $L$ encodes the $T_k(Ah)$ and $S_k(Ah)$ matrices, as indicated in Eqs. (\ref{eq:T_k}-\ref{eq:matrix_L}). As the matrix $A$ enters these two matrix functions, the error is multiplied by $h$ and accompanied by another error originating from truncating the matrix functions at level $k$. The latter is already contained in $\delta'$, hence we do not include it in the below equation. The nonzero entries of $L$ therefore carry a grid discretization error of
\begin{align}
\mathcal{O}\left(h[\Delta x^2+\Delta v^2]\label{eq:error_at_h_quantum}
\right).
\end{align}
Note that this is analogous to the same error on the classical solution at time $t=h$ given in Eq. \eqref{eq:classical_error_h}. This is then amplified by the condition number of $L$, 
which is bounded by \cite[Theorem 4]{krovi}
\begin{align}
\kappa(L)\leq (m+p)C(A)(1+\delta)e(1+e),
\end{align}
where 
\begin{align}
C(A) = \sup_{t\in[0,T]}\norm{\exp(At)}
\end{align}
is a measure of the growth in the state vector $z$ induced by the matrix $A$. According to \cite[Lemma 16]{krovi}, we have $C(A)\leq 1$ after the rescaling described in Subsection \ref{rescaling}. Additionally, we have $m=p=T/h$ according to Eq. \eqref{eq:parameters_chosen}, hence $\kappa(L)=\mathcal{O}(T/h)$. Note how multiplying Eq. \eqref{eq:error_at_h_quantum} by $\kappa(L)$ is entirely analogous to the temporal accumulation of the error in the fully classical case (when going from Eq. \eqref{eq:classical_error_h} to Eq. \eqref{eq:eps(T)}). This all means that the error originating from phase space discretization within the quantum algorithm scales as
\begin{align}
\varepsilon_c=\mathcal{O}\left(\frac{T}{N_x^2+N_v^2}\right).\label{eq:quantum_classical_error}
\end{align}

The other classical errors, $\delta$ and $\delta'$ are set according to the size of $\varepsilon_q$, as discussed in Subsection \ref{errors_section}. This way we may write the total error on the entries of the state $|y_{1,m}\rangle$ as $\mathcal{O}(\varepsilon_q+\varepsilon_c)$.

\section{Complexity analysis}\label{complexity_section}

In the complexity analysis, we reparameterize the grid size by the single quantity $N = N_x N_v$, with $N_x$ and $N_v$ assumed to scale comparably; equivalently, $N_x$ and $N_v$ are both of order $N^{1/2}$.

We now present the query- and gate complexities for this fixed-aspect-ratio grid. The complexities are obtained from Eqs.~\eqref{eq:querycomplexity} and \eqref{eq:gatecomplexity_factor} by inserting the Vlasov-equation-specific bounds for $s_A$, $\|A\|$, $N_C(\delta)$, $g_u$, and $\Omega(\delta',\delta)$ derived in Appendix~\ref{appendix:quantities_for_complexity}, as shown in detail in Appendix~\ref{sec:compl_in_terms_of_Nx_Nv_N}. Thus, the expressions below are the generic complexity bounds from Subsection~\ref{gatecomplexity} specialized to our discretized Vlasov system.

A key input to the complexity is the sparsity of the Carleman matrix. The primitive sparsities are $N$ for $\bar{F}^{(0)}$, $5$ for $F^{(1)}$, and $2N$ for $\bar{F}^{(2)}$, hence $s=2N=\mathcal{O}(N)$ and $s_A=3N_C(\delta)s$. Although the underlying QLSA depends only polylogarithmically on the dimension of the linear system, the factors $s_A$ and $N_C(\delta)=\mathcal{O}((N_x+N_v^{1/3})\log(TN_v^2/\delta))$ enter outside these dimension-dependent polylogarithms. Therefore, the full algorithm is not polylogarithmic in the grid size.

Now for the truncation errors we choose
\begin{align}
\begin{split}
\delta&=\frac{\varepsilon_q}{4}=\mathcal{O}(\varepsilon_q),\\
\delta'&=\frac{\varepsilon_q}{4+\varepsilon_q} \frac{1}{\sqrt{N_C(\delta=\varepsilon_q/4)}}
=\mathcal{O}\left(
\varepsilon_q \left[\frac{\log(1/\norm{\Bar{u}_{\text{in}}})}{\log(T\norm{\Bar{F}^{(2)}}/\varepsilon_q\norm{\Bar{u}(T)})}\right]^{1/2}
\right),
\end{split}
\end{align}
which ensures asymptotically that $\delta+(1+\delta)\delta'\sqrt{N_C(\delta)}= \varepsilon_q/2$, in accordance with Section~\ref{errors_section}. As for the phase space discretization error, we set $N=\mathcal{O}(T/\varepsilon_c)$ according to Eq. \eqref{eq:quantum_classical_error}. Additionally, whenever the parameter $\nu_0$ appears, we substitute $\nu_0\geq \mathcal{O}(N_v^{3/2})=\mathcal{O}(N^{3/4})$ from Eq. \eqref{eq:nu_0_scaling} to stay within the boundary of convergence, i.e. to keep $R\lesssim 1$.

To compare the classical and quantum algorithms on a meaningful basis, we express the complexity in terms of the total error rather than solely in terms of the grid size. This is because, for a fixed discretization level $N$, the quantum and classical schemes exhibit different scaling behavior with respect to the discretization and algorithmic error contributions. Therefore, the relevant figure of merit is the achieved accuracy, which combines both discretization and algorithmic errors in a unified way.

Defining the parameter $\eta\equiv T/(\varepsilon_q\varepsilon_c)$, the above procedure leads to a query complexity of\,\footnote{We use the identity $\log(X^aY^b)=\mathcal{O}(\log[XY])$ with some fixed $a,b>0$ for $X,Y\to\infty$, as well as $\log[X\log(X)]=\mathcal{O}(\log[X])$ for $X\to\infty$ to simplify our expressions.}
\begin{align}
\begin{split}
\mathcal{O}\Bigg(&T^{9/2}\varepsilon_c^{-7/2}\frac{ \log^{\frac{5}{2}}(\eta)}{\log [\log (\eta)]}\text{poly}\Bigg[\frac{T^{1/2}}{\varepsilon_c^{1/2}}\log(\eta)\Bigg]
\text{polylog}\Bigg[
\log(\eta),
\frac{T}{\varepsilon_c}\log(\eta),\frac{T}{\varepsilon_c},\frac{1}{\varepsilon_q}
\Bigg]\Bigg)\\
=\mathcal{O}\Bigg(&T^{9/2}\varepsilon_c^{-7/2}\frac{ \log^{\frac{5}{2}}(\eta)}{\log [\log (\eta)]}\text{poly}\Bigg[\frac{T}{\varepsilon_c}\log(\eta)\Bigg]
\text{polylog}\Bigg[\frac{T}{\varepsilon_c},\frac{1}{\varepsilon_q}
\Bigg]\Bigg),
\label{eq:T_eps_querycomplexity_final}
\end{split}
\end{align}
and a gate complexity which is larger than the above by the factor
\begin{align}
\begin{split}
&\mathcal{O}\left(
\text{polylog}\left[
\log(\eta),\frac{T}{\varepsilon_c}\log(\eta),\frac{1}{\varepsilon_q}
\right]
\right)\\
=&\mathcal{O}\left(
\text{polylog}\left[
\frac{T}{\varepsilon_c},\frac{1}{\varepsilon_q}
\right]
\right),
\end{split}\label{eq:T_eps_gatecomplexity_final}
\end{align}
where, in the second lines of Eqs.~(\ref{eq:T_eps_querycomplexity_final}-\ref{eq:T_eps_gatecomplexity_final}), we have suppressed asymptotically redundant factors that do not affect the leading-order scaling behavior.
The leading order query- and gate complexities without logarithmic terms read
\begin{align}
\widetilde{\mathcal{O}}\left(T^{9/2}\varepsilon_c^{-7/2} \text{poly}\left[T\varepsilon_c^{-1}\right]
\right).\label{eq:leading_complexity}
\end{align}
Equation~\eqref{eq:leading_complexity} shows that the polylogarithmic QLSA dependence on dimension is outweighed here by Vlasov-specific factors outside the polylogarithms, namely the sparsity, $\|A\|$, and $N_C(\delta)$. Using $N=\mathcal{O}(T/\varepsilon_c)$ then gives polynomial dependence on $T$ and $\varepsilon_c^{-1}$.
This may now be compared to the time complexity of the classical solution of the same finite difference equation. In Appendix \ref{appendix:classical_error_complexity} we show that to be 
\begin{align}
\mathcal{O}\left(T^{2} \varepsilon_c^{-1}\right),\label{eq:classical_complexity_again}
\end{align}
where the error $\varepsilon_c$ is completely analogous to the classical error in the quantum complexities that is denoted the same. We conclude that regardless of the exact form the unknown polynomial takes, the query and gate complexities in Eqs.~(\ref{eq:T_eps_querycomplexity_final}--\ref{eq:T_eps_gatecomplexity_final}) are polynomially larger than the classical time complexity in Eq.~\eqref{eq:classical_complexity_again}. This polynomial overhead arises from the Vlasov-specific growth of the Carleman matrix size, sparsity, and number of linearization levels under the present mapping and discretization.

\section{Coupling to Ampere's law}\label{sec:ampere}
In this section we explore why coupling the Vlasov equation to Ampere's law would lead to $\mu(F^{(1)})\geq 0$, which would cause Carleman linearization to not converge.

\subsection{Finite difference equations}\label{sec:finite_diff_eq_ampere}
Discretizing Eq. \eqref{eq:ampereevolution} using the phase space grid described in Section \ref{sec:coord_discretization} gives the finite difference equations
\begin{subequations}
\begin{align}
\frac{d}{dt}f_{ij} = &
+\frac{q}{m_e}E_i \left. \frac{\partial}{\partial v} \right|_{ij} f
&& \text{(quadratic)} \label{eq:1species_discretized_a_ampere} \\
& -v_j \left. \frac{\partial}{\partial x} \right|_{ij} f 
&& \text{(linear)} \label{eq:1species_discretized_b_ampere} \\
& -\nu(v_j)f_{ij}
&& \text{(linear)} \label{eq:1species_discretized_c_ampere} \\
& +\nu(v_j)f^{M}_j
&& \text{(inhomogeneous)}, \label{eq:1species_discretized_d_ampere} \\
\frac{d}{dt}E_{i} = & +\frac{q}{\varepsilon_0} \int v_J f_{iJ}\, dv_J
&& \text{(linear)} \label{eq:1species_discretized_e_ampere}\\
& -\frac{q}{\varepsilon_0} \int v_J f^{\text{bg}}_{J}\, dv_J
&& \text{(inhomogeneous)}, \label{eq:1species_discretized_f_ampere}
\end{align}
\end{subequations}
where $E_i(t)=E(x_i,t)$ is understood. The contents of the $\mathcal{O}(E^1f^1)$ line encode the nonlinearity, hence they determine $F^{(2)}$. Similarly, the  $\mathcal{O}(f^1)$ lines determine $F^{(1)}$ and the $\mathcal{O}(f^0)$ ones determine $F^{(0)}$.

When coupling to Ampere's law, the electric field is treated as a dynamical object similarly to the distribution matrix. The state vector is chosen to be formulated by appending the $E_i$'s at the end of $\text{vec}(f)$ as
\begin{align}
u= \left[ f_{1,1}, \:  \cdots, \: f_{1,N_v}, \: f_{2,1}, \: \cdots, \: f_{2,N_v},\: \dots, \: f_{N_x,N_v},\: E_1 ,\dots ,\:E_{N_x} \right]^\mathsf{T}.
\label{eq:amperestatevector}
\end{align}
We provide the conversion formulas for this vectorization in Appendix~\ref{appendix:vectorization_of_u_ampere}.

\subsection{Convergence}\label{subsec:Ampere_conv}

The fundamental problem with coupling to Ampere's law is that $\mu(F^{(1)})<0$ is not satisfied. To see this, let us write the \emph{linear} part of the time evolution from Eq. \eqref{eq:ampereevolution} in matrix notation as 
\begin{align}
\frac{\partial }{\partial t}
    \left[\begin{array}{c}
    f\\[8 pt]
    E\\
    \end{array}\right]
    =
    \begin{pmatrix}
    -v\frac{\partial}{\partial x}-\nu(v)  &0 \: \\[8 pt]
    \frac{q}{\varepsilon_0}\int v dv  &0 \:\\
    \end{pmatrix}
   \left[\begin{array}{c}
    f\\[8 pt]
    E\\
    \end{array}\right]+\text{quadratic and inhomogeneous terms.} \label{eq:linear_part_of_ampere}
\end{align}
The role of Eq.~\eqref{eq:linear_part_of_ampere} is only to identify the block structure of the linearized evolution. In the Ampere formulation, the electric field enters the Vlasov equation through the Lorentz-force term $E\,\partial f/\partial v$, which is quadratic in the state variables once both $f$ and $E$ are included in $u$. Hence this contribution belongs to $F^{(2)}$, not to $F^{(1)}$. Moreover, Ampere's law contains a term linear in $f$, but no term linear in $E$. Therefore the columns of $F^{(1)}$ corresponding to the $E$ variables are identically zero, also meaning that $F^{(1)}$ will have vanishing eigenvalues for any discretization scheme used.

For the specific discretization used here, schematically written in Eqs.~(\ref{eq:1species_discretized_a_ampere}-\ref{eq:1species_discretized_f_ampere}) and visualized in Fig.~\ref{fig:double_F1}(b), there are $N_x$ such electric-field variables. Thus $F^{(1)}$ has at least $N_x$ zero columns. Equivalently, the subspace spanned by vectors supported only on the electric-field components is contained in the kernel of $F^{(1)}$. Hence $F^{(1)}$ has at least $N_x$ zero eigenvalues, and therefore $\alpha(F^{(1)})\geq 0$, where $\alpha(\cdot)$ is defined in Appendix~\ref{appendix:norms}. Now Eq.~\eqref{eq:evaluerelation} tells us that
\begin{align}
0\leq \alpha\left(F^{(1)}\right)\leq \mu\left(F^{(1)}\right),
\end{align}
and consequently the condition $\mu(F^{(1)})<0$ cannot be satisfied. The quantum algorithm therefore does not converge for the Ampere-coupled formulation, regardless of the values of the parameters in Eqs.~(\ref{eq:1species_discretized_a_ampere}-\ref{eq:1species_discretized_f_ampere}).

\section{Discussion}\label{section:discussion}

In our procedure, the value of the $R$ parameter only depends on $N_v$ and not $N_x$ in the asymptotic limit. This is a result of the difference between the scalings of the spectral- and Frobenius norms of $F^{(2)}$, as stated in Appendix \ref{appendix:F2_norm}. $R$ depending on $N_v$ only means that if our physical system has some base-line collision frequency $\nu_0$, then it puts a restriction on the maximum value $N_v$ may take but puts no restriction on $N_x$. When using a physical model for $\nu_0$, we may formulate that as an inequality between $N_v$, temperature $\mathcal{T}$ and position periodicity $x_{\max}$, as shown in Eq. \eqref{eq:N_v_restriction}. 
This is qualitatively a similar restriction as the well-known CFL condition in numerical analysis that connects computational and physical parameters \cite{numerical}. We also would like to note that our formula for $R$, as given in Eq. \eqref{eq:nu_0_scaling}, strongly resembles the same formula derived for the Carleman linearization of the Navier-Stokes equations in Ref. \cite{linearization_efficiency}. In that system, linear dissipativity is provided by viscosity, that is qualitatively analogous to the collisions in plasma.

Eq. \eqref{eq:parameters_chosen} suggests that the number of required Carleman steps $N_C$ is  logarithmic. However, $1/\norm{\Bar{u}_{\text{in}}}$ asymptotically approaches $1$ in the denominator and hence causing the expression to diverge. This is due to the fact that $\gamma $ approaches $\norm{u_{\text{in}}}$ asymptotically. To resolve the issue, in Appendix \ref{section:Carleman_steps} the second order contributions were collected and the logarithm was Taylor expanded with respect to them. Therefore, $N_C$ is actually polynomial instead of logarithmic with respect to system size. Consequently, $d$, the dimension of the Carleman linearized system from Eq. \eqref{eq:carlemanmatrix}, is exponential in $N_x$ and $N_v$. As a result of the query- and gate complexities of the underlying QLSA being polylogarithmic in $d$, the final complexities of our algorithm gain a polynomial factor in $N_x$ and $N_v$, which is then converted into a polynomial factor in $T$ and $\varepsilon_c$. This could be hypothetically resolved by using a QLSA whose complexities scale as $\log\, \log\, d$. This is highly unlikely, as the Hilbert space can only store an exponential amount of information with respect qubit number, instead of super-exponential.

Another large addition to the complexity of the problem is the sparsity of $A$ that inherits the maximum of the sparsities of the $F^{(k)}$ matrices. Since $F^{(0)}$ is dense due to the Maxwellian in the Krook collision operator and $F^{(2)}$ encodes two double integrals in each row, they have $s=N,\: 2N$, respectively. On the other hand, $F^{(1)}$ encodes finite difference derivatives and its sparsity is fixed to five. As a result, the sparsity of the linearized system grows linearly with system size, making it dense. Note that the sparsity of $F^{(0)}$ is the consequence of our choice for the collision operator but the sparsity of $F^{(2)}$ is inherently part of the system, due to the inversion of Gauss's law. The factor the sparsity adds to the complexity could be potentially reduced by using a QLSA that is optimized with respect to sparsity \cite{dense}. 

Furthermore, as it was mentioned in Subsection \ref{sec:ingredients_subsec}, the choice for the initial condition $u_{\text{in}}$ of the system is heavily restricted. This is due to the requirement that if one wishes to control the value of $R$ in the algorithm with the collision frequency $\nu_0$, the terms independent of $\nu_0$ must vanish in the asymptotic limit. This is a consequence of the mathematical form of the collision operator. While the Krook one contains an inhomogeneous term, others such as the Fokker-Planck do not. In those cases the restriction on $u_{\text{in}}$ is expected to be significantly looser, since then $F^{(0)}=0$. On the contrary, the Fokker-Planck operator would make $F^{(2)}$ significantly more complicated if multiple species are treated as dynamic.

The output state of the algorithm encodes the full phase-space distribution $f(T)$ through the level-1 block $|y_{1,m}\rangle$. In practice, extracting a classical prediction would require a problem-specific measurement protocol, which we do not develop here. The most natural quantities are low-dimensional observables obtained as linear or quadratic functionals of $f$, such as spatial density moments, velocity moments, current-like integrals, or other coarse-grained diagnostics relevant to the plasma model. Such quantities can in principle be estimated from $|y_{1,m}\rangle$ by measuring appropriate observables on the third register, possibly after additional amplitude estimation or state-preparation steps. Since the present work is concerned with the mapping and its complexity, we leave a detailed analysis of measurement strategies and observable extraction to future work.

It is important to discuss whether the choice of vectorization affects the convergence of the quantum algorithm. In Section \ref{sec:mapping} we stated that our choice is row-major vectorization, but many other choices exist that are equally valid from a mathematical point of view, e.g. column-major vectorization. The output of different vectorizations of the same matrix (or multidimensional array in general) are permutations of each other. This means that if $u=\text{vec}_1(f)$ and $\Tilde{u}=\text{vec}_2(f)$ are different vectorizations of $f$, then there exists a permutation matrix (also known as commutation matrix) $P$ such that $u=P\Tilde{u}$. The matrix $P$ has exactly one $1$ in each row and column, with $0$'s everywhere else. Furthermore, it is orthogonal in the sense that $P^\mathsf{T}P=PP^\mathsf{T}=\mathbb{I}$. Substituting $u=P\Tilde{u}$ into Eq. \eqref{eq:krovidef} and multiplying by $P^\mathsf{T}$, we obtain
\begin{align}
\frac{d\Tilde{u}}{dt} = \Tilde{F}^{(2)}\left(\Tilde{u}\otimes \Tilde{u}\right) +\Tilde{F}^{(1)} \Tilde{u} +\Tilde{F}^{(0)},
\end{align}
where the permuted matrices are
\begin{align}
\Tilde{F}^{(2)} = P^\mathsf{T} F^{(2)} \left(P\otimes P\right), \quad\quad \Tilde{F}^{(1)} = P^\mathsf{T} F^{(1)} P,\quad\quad \Tilde{F}^{(0)} = P^\mathsf{T} F^{(0)}.\label{eq:tilde_matrices}
\end{align}
The computational complexity of the problem is formulated in terms of the norms of the above matrices, meaning that as long as unitarily/orthogonally invariant norms are used, the complexity is invariant under the choice of vectorization and is an intrinsic property of the finite difference scheme. The spectral norm used in this paper, as well as other norms characterized by singular values only\footnote{Examples besides the spectral norm are the Frobenius norm and the trace/nuclear norm. Counterexamples are the $1$- and $\infty$ norms.}, satisfy this property \cite{matrix_book}. For these norms, $\norm{UAV}=\norm{A}$ holds for all unitary/orthogonal $U,V$, which means that $\norm{\Tilde{F}^{(k)}}=\norm{F^{(k)}}$ holds in Eq. \eqref{eq:tilde_matrices} for $k=0,1,2$, leaving the complexity invariant. Upon substituting $F^{(1)}= P \Tilde{F}^{(1)} P^\mathsf{T}$ into the eigenvalue equation satisfied by it, we also see that the permutation leaves the eigenvalues invariant. This is also true for the eigenvalues of the symmetric part of $F^{(1)}$, i.e. the log norm $\mu(F^{(1)})$ is invariant as well.

The above argument holds regardless of the dimension of the original array to vectorized, i.e. when vectorizing the six dimensional array containing the values of $f_{i_1,i_2,i_3,j_1,j_2,j_3}$ that arises when discretizing the distribution function in 3+3 dimensional phase space, any output vector is the permutation of any other possible output.

Additionally, it is worth noting that the $R<1$ requirement is sufficient but not necessary, as explored in Ref. \cite{liu}. They performed numerical simulations concerning the convergence of Carleman linearization for Burger's equation with $R\approx 44$ and found that the error still decreases exponentially as $N_C$ is increased. We anticipate that this conclusion would also hold for the Vlasov equation, meaning that the restrictions on the plasma parameters discussed in Subsection \ref{sec:connection_to_plasma} could be relaxed. 

Last but not least, a recent refinement of Carleman stability analysis by Costa et al. suggests that the convergence criteria can exhibit a more favorable structure when formulated using rescaling techniques (that are similar in nature to the one used here via $\gamma$, and in Refs. \cite{krovi,liu}) and more importantly, the $l_\infty$-norm rather than the $l_2$-norm \cite{Costa_2025}. In particular, their analysis indicates that certain apparent dependencies on discretization size may be weakened when the nonlinear evolution is sufficiently well-conditioned and homogeneous. However, these improvements rely on a setting without external forcing or inhomogeneous driving terms. In the present work, the Vlasov–Carleman system contains a nontrivial inhomogeneous contribution $F^{(0)}$ originating from the collision-induced relaxation toward the Maxwellian, which plays a central role in the convergence balance. As a result, the direct applicability of the above framework is limited, although their results strongly suggest that alternative norm choices may lead to quantitatively different (but structurally similar) stability thresholds. Exploring such norm-dependent refinements in the presence of inhomogeneous driving would be an interesting direction for future work.

\section{Conclusion and prospects}\label{section:conclusion}
The central result of this work is that the convergence requirements of the Carleman-linearization quantum algorithm severely restrict its applicability to physically relevant plasma simulations. By mapping the nonlinear Vlasov equation onto the framework of Ref.~\cite{krovi}, we were able to analyze these requirements quantitatively. For the Vlasov-Gauss system, convergence can only be achieved when the collision frequency is much larger than physically realistic values for practical grid sizes. For the Vlasov-Ampere system, the convergence conditions are not satisfied within our formulation because the linear part of the evolution is not dissipative.

We showed for the first time how the nonlinear Vlasov equation can be formulated within the Carleman-linearization framework of Ref.~\cite{krovi}. In Section~\ref{vlasov_section} we derived the finite-difference discretization of the Vlasov equation and in Section~\ref{sec:mapping} we mapped the resulting system onto the quadratic ODE form required by the quantum algorithm, Eq.~\eqref{eq:krovidef}. This mapping enabled a detailed analysis of the convergence conditions and computational complexity of the algorithm when applied to plasma simulations.

The query- and gate complexities of the problem were analyzed with the Gauss's law coupling and with the assumption that the convergence criteria are met. It was found that their upper bounds of both are polynomially larger than the time complexity of the most straightforward classical algorithm that solves the same finite difference equation. The largest contributors to the quantum complexities are $\norm{A}$, the norm of the Carleman linearized evolution matrix; the sparsity of $A$ that is linear in the size of the phase-space grid; and finally $N_C$, the number of Carleman linearization steps which grows polynomially with system size as well.

Jennings et al. \cite{jennings2025quantumalgorithmsgeneralnonlinear} have recently improved on the approach of Ref.~\cite{krovi}, and adopting techniques of Ref.~\cite{jennings2025quantumalgorithmsgeneralnonlinear} may change the present conclusions. Specifically, Ref.~\cite{jennings2025quantumalgorithmsgeneralnonlinear} generalizes the convergence analysis of the Carleman embedding technique previously used in the literature for multiple classes of ODE systems. In particular, for stable systems with $\alpha(F^{(1)})<0$, such as the one derived in this paper from the Vlasov--Gauss system, \cite[Theorem~3.6]{jennings2025quantumalgorithmsgeneralnonlinear} gives a family of convergence conditions induced by a positive-definite matrix $P>0$ satisfying the Lyapunov inequality
\begin{align}
PF^{(1)}+\left(F^{(1)}\right)^\dagger P<0.
\end{align}
The matrix $P$ defines a weighted inner product $\langle x,y\rangle_P:=x^\dagger P y$ for vectors $x$ and $y$, with associated norm $\|x\|_P:=\sqrt{x^\dagger P x}$, and corresponding induced operator norms 
\begin{align}
\begin{split}
\norm{F^{(2)}}_P&=\norm{P^{1/2}F^{(2)}\left(P^{-1/2}\otimes P^{-1/2}\right)}, \\
\mu_P(F^{(1)})&:=\max_{x\neq 0}\operatorname{Re}\frac{\langle F^{(1)}x,x\rangle_P}{\langle x,x\rangle_P}<0.
\end{split}
\end{align}
In this setting, the convergence condition is expressed in terms of the generalized parameter
\begin{align}
R_P=\frac{1}{-\mu_P(F^{(1)})}\left(\|F^{(2)}\|_P \|u_{\rm in}\|_P+\frac{\|F^{(0)}\|_P}{\|u_{\rm in}\|_P}\right).
\end{align}
Optimizing over the choice of $P$ may therefore relax the effective convergence requirements, including the lower bound on the collision frequency, and may potentially change the quantitative conclusions of the present analysis. We leave such an optimization to future work.

This direction has since been pursued by Christensen et al.~\cite{christensen2026improvedconvergencecarlemanembeddedquantum}, who construct a Lyapunov-based convergence analysis for the Vlasov–Poisson system using a Fourier–Hermite expansion of the distribution function shifted by the background Maxwellian, numerically optimizing the associated Lyapunov matrix. They obtain convergence guarantees for physically reasonable collision frequencies in some regimes, directly addressing the restriction identified in Subsection~\ref{sec:connection_to_plasma} for the formulation considered here.

We also note that Ref.~\cite{jennings2025quantumalgorithmsgeneralnonlinear} extends the convergence analysis to certain conservative systems, where the spectrum of the linear part may contain eigenvalues with zero real part. This may be relevant for the Ampere--Vlasov formulation discussed in Section~\ref{sec:ampere}, which was not pursued here because the previously available convergence guarantees did not apply in the presence of zero eigenvalues. Revisiting this formulation under the more recent conservative-system conditions of Ref.~\cite{jennings2025quantumalgorithmsgeneralnonlinear} is therefore an interesting direction for future work.

A related but distinct approach has been developed by Berntson et al.~\cite{berntson2026endtoendquantumalgorithmweakly}, who present an end-to-end quantum algorithm for a three-dimensional, weakly nonlinear electron–ion plasma model with Debye screening and Krook relaxation. Their construction combines a Lyapunov/free-energy transformation with Carleman linearization, hierarchical block-encoding of the dense field-interaction terms, and an explicit observable-extraction procedure. To date this is the most complete end-to-end treatment of a nonlinear plasma model in this framework, however, by the authors' own account, their convergence guarantees currently apply only to nonlinearities well below the strength typically of physical interest.

The development of quantum algorithms with the same input formalism as Eq. (\ref{eq:krovidef}) is an active area of research, partially because higher dimensional polynomial nonlinearities can be transformed into quadratic ones \cite{forets2017explicit}. Hence it is likely that quantum algorithms with less strict constraints and lower complexities will be put forward, and then our mapping will be applicable to those without major modifications required. An example would be applying QLSA's that are optimized with respect to system dimension, or sparsity.

Future work could be in multiple directions. As it may be seen from the complexities, the dependence on the error originating from phase space discretization is polynomial, meanwhile others are only logarithmic. This might be mitigated by solving the PDE directly on a quantum device rather than discretizing the PDE into an ODE system and solving the latter. Performing this would require an embedding method for nonlinear PDE systems, as a generalization of those for ODE ones (i.e a generalization of Carleman linearization or homotopy perturbation). A prototype for such a mathematical framework was recently proposed by us in Ref. \cite{arxiv_carleman}. However, it is yet unclear whether such a high-dimensional family of linear PDEs can be efficiently and accurately solved on analog quantum simulators operating with continuous variables. If that can be achieved, the error dependence of the complexity of our algorithm would be relaxed from polynomial to logarithmic. Furthermore, we note that prior work on the wave equation indicates that utilizing higher order discretizations of differential operators results in more advantageous quantum complexity scalings, similarly how they manifest in classical algorithms \cite{wave}.

Additionally, lattice-Boltzmann-based quantum methods may be an interesting direction for future work in plasma physics due to their use in related nonlinear fluid-mechanics models.

Another direction could be  to generalize our procedure to the full (3+3) dimensional Vlasov-Maxwell system. However, since that is formulated in terms of integrals of the phase space distribution function within the past lightcones (in terms of retarded time), the system is actually made of  \emph{delay} differential equations. It can be put into normal (integro-) differential equation form by working in the non-relativistic limit. By carrying this out, one might find that the linear part of the evolution not being dissipative when coupling to Ampere's/Faraday's law is just a consequence of working in one dimension and having no magnetic field. It would also be insightful to know how the complexity of the problem varies with dimension.

Furthermore, implementing more complex and physically realistic collision operators remains an important direction for future work. That might change the structure of the $F^{(i)}$ matrices in a way that brings the convergence with $R<1$ closer to physically meaningful scenarios.

\section{Acknowledgments}
TV and AD were supported, in part, by the UK Research and Innovation Exascale Computing ALgorithms \& Infrastructures Benefiting UK Research (ExCALIBUR) project Quantum Enhanced Verified Exascale Computing (QEVEC, EP/W00772X/2). AD is also supported, in part, by the  Hub for Quantum Computing via Integrated and Interconnected Implementations (QCI3) (EP/Z53318X/1), and TV by the Charles Coulson Scholarship at the Mathematical Institute, University of Oxford.

TV also thanks Bálint Koczor for helpful guidance and discussions regarding the resubmission of this manuscript, and David Jennings for suggesting the research internship at the University of Warwick.

\section{Author Contributions}
TV carried out the main research, including the analytical development, complexity analysis, and preparation of the manuscript. BA supervised the original project at Imperial College London and contributed to the plasma-physics aspects of the work. AD and TG contributed to the development and interpretation of the work through discussions and feedback, with AD providing particular input on the quantum-computing aspects and TG on the plasma-physics aspects. All authors reviewed and approved the final manuscript. A Large Language Model (ChatGPT) was used to assist with language editing, and discussing the presentation and organization of the manuscript. All authors take full responsibility for the content of the work.

\bibliographystyle{plainnat}
\bibliography{ref}

\begin{thebibliography}{66}
\providecommand{\natexlab}[1]{#1}
\providecommand{\url}[1]{\texttt{#1}}
\expandafter\ifx\csname urlstyle\endcsname\relax
  \providecommand{\doi}[1]{doi: #1}\else
  \providecommand{\doi}{doi: \begingroup \urlstyle{rm}\Url}\fi

\bibitem[Alekseenko and Euler(2013)]{krook4}
A.~Alekseenko and Craig Euler.
\newblock A {B}hatnagar–{G}ross–{K}rook kinetic model with velocity-dependent collision frequency and corrected relaxation of moments.
\newblock \emph{Continuum Mechanics and Thermodynamics}, 01 2013.
\newblock \doi{10.1007/s00161-014-0407-0}.

\bibitem[Ameri et~al.(2023)Ameri, Ye, Cappellaro, Krovi, and Loureiro]{plasmaquant2}
Abtin Ameri, Erika Ye, Paola Cappellaro, Hari Krovi, and Nuno~F. Loureiro.
\newblock Quantum algorithm for the linear {V}lasov equation with collisions.
\newblock \emph{Phys. Rev. A}, 107:\penalty0 062412, Jun 2023.
\newblock \doi{10.1103/PhysRevA.107.062412}.

\bibitem[An et~al.(2025)An, Liu, Wang, and Zhao]{an2023theoryquantumdifferentialequation}
Dong An, Jin-Peng Liu, Daochen Wang, and Qi~Zhao.
\newblock {Quantum Differential Equation Solvers: Limitations and Fast-Forwarding}.
\newblock \emph{Communications in Mathematical Physics}, 406\penalty0 (8):\penalty0 189, 2025.
\newblock \doi{10.1007/s00220-025-05358-7}.

\bibitem[Arber and Vann(2002)]{vlasov_solver_1}
T.D. Arber and R.G.L. Vann.
\newblock {A Critical Comparison of Eulerian-Grid-Based Vlasov Solvers}.
\newblock \emph{Journal of Computational Physics}, 180\penalty0 (1):\penalty0 339--357, 2002.
\newblock ISSN 0021-9991.
\newblock \doi{10.1006/jcph.2002.7098}.

\bibitem[Berntson et~al.(2026)Berntson, Jennings, Lostaglio, and Parker]{berntson2026endtoendquantumalgorithmweakly}
Bjorn~K. Berntson, David Jennings, Matteo Lostaglio, and Scott Parker.
\newblock {An end-to-end quantum algorithm for weakly nonlinear plasma physics with superquadratic speedup}.
\newblock 2026.
\newblock \doi{10.48550/arXiv.2607.14308}.

\bibitem[Berry(2014)]{berry1}
Dominic~W Berry.
\newblock High-order quantum algorithm for solving linear differential equations.
\newblock \emph{Journal of Physics A: Mathematical and Theoretical}, 47\penalty0 (10):\penalty0 105301, 2014.
\newblock \doi{10.1088/1751-8113/47/10/105301}.

\bibitem[Berry et~al.(2017)Berry, Childs, Ostrander, and Wang]{berry2}
Dominic~W. Berry, Andrew~M. Childs, Aaron Ostrander, and Guoming Wang.
\newblock {Quantum Algorithm for Linear Differential Equations with Exponentially Improved Dependence on Precision}.
\newblock \emph{Communications in Mathematical Physics}, 356\penalty0 (3):\penalty0 1057--1081, 2017.
\newblock \doi{10.1007/s00220-017-3002-y}.

\bibitem[Bhatnagar et~al.(1954)Bhatnagar, Gross, and Krook]{krook1}
P.~L. Bhatnagar, E.~P. Gross, and M.~Krook.
\newblock {A Model for Collision Processes in Gases. {I}. Small Amplitude Processes in Charged and Neutral One-Component Systems}.
\newblock \emph{Phys. Rev.}, 94:\penalty0 511--525, May 1954.
\newblock \doi{10.1103/PhysRev.94.511}.

\bibitem[Brustle and Wiebe(2025)]{Brustle2025quantumclassical}
Noah Brustle and Nathan Wiebe.
\newblock {Quantum and classical algorithms for nonlinear unitary dynamics}.
\newblock \emph{{Quantum}}, 9:\penalty0 1741, May 2025.
\newblock ISSN 2521-327X.
\newblock \doi{10.22331/q-2025-05-13-1741}.

\bibitem[Carleman(1932)]{carleman0}
Torsten Carleman.
\newblock {Application de la théorie des équations intégrales linéaires aux systèmes d'équations différentielles non linéaires}.
\newblock \emph{Acta Mathematica}, 59\penalty0 (none):\penalty0 63 -- 87, 1932.
\newblock \doi{10.1007/BF02546499}.

\bibitem[Chankin et~al.(2012)Chankin, Coster, and Meisl]{vlasov_solver_2}
A.V. Chankin, D.P. Coster, and G.~Meisl.
\newblock {Development and Benchmarking of a New Kinetic Code for Plasma Periphery (KIPP)}.
\newblock \emph{Contributions to Plasma Physics}, 52\penalty0 (5-6):\penalty0 500--504, 2012.
\newblock \doi{10.1002/ctpp.201210039}.

\bibitem[Chen(2016)]{plasma2}
Francis~F Chen.
\newblock \emph{{Introduction to plasma physics and controlled fusion}}, volume~1.
\newblock Springer, 2016.
\newblock \doi{10.1007/978-3-319-22309-4}.
\newblock Third Edition.

\bibitem[Childs and Liu(2020)]{childs}
Andrew~M. Childs and Jin-Peng Liu.
\newblock {Quantum Spectral Methods for Differential Equations}.
\newblock \emph{Communications in Mathematical Physics}, 375\penalty0 (2):\penalty0 1427--1457, 2020.
\newblock \doi{10.1007/s00220-020-03699-z}.

\bibitem[Childs et~al.(2021)Childs, Liu, and Ostrander]{partial}
Andrew~M Childs, Jin-Peng Liu, and Aaron Ostrander.
\newblock High-precision quantum algorithms for partial differential equations.
\newblock \emph{Quantum}, 5:\penalty0 574, 2021.
\newblock \doi{10.22331/q-2021-11-10-574}.

\bibitem[Christensen et~al.(2026)Christensen, Goffrey, and Datta]{christensen2026improvedconvergencecarlemanembeddedquantum}
Matthew Christensen, Tom Goffrey, and Animesh Datta.
\newblock {Improved Convergence of Carleman-Embedded Quantum Algorithm for the Vlasov-Poisson System}.
\newblock 2026.
\newblock \doi{10.48550/arXiv.2607.28426}.

\bibitem[Costa et~al.(2019)Costa, Jordan, and Ostrander]{wave}
Pedro C.~S. Costa, Stephen Jordan, and Aaron Ostrander.
\newblock Quantum algorithm for simulating the wave equation.
\newblock \emph{Phys. Rev. A}, 99:\penalty0 012323, Jan 2019.
\newblock \doi{10.1103/PhysRevA.99.012323}.

\bibitem[Costa et~al.(2025)Costa, Schleich, Morales, and Berry]{Costa_2025}
Pedro C.~S. Costa, Philipp Schleich, Mauro E.~S. Morales, and Dominic~W. Berry.
\newblock Further improving quantum algorithms for nonlinear differential equations via higher-order methods and rescaling.
\newblock \emph{npj Quantum Information}, 11\penalty0 (1), August 2025.
\newblock ISSN 2056-6387.
\newblock \doi{10.1038/s41534-025-01084-z}.

\bibitem[Dawson(1983)]{vlasov_solver_4}
John~M. Dawson.
\newblock Particle simulation of plasmas.
\newblock \emph{Rev. Mod. Phys.}, 55:\penalty0 403--447, Apr 1983.
\newblock \doi{10.1103/RevModPhys.55.403}.

\bibitem[Demirdjian et~al.(2026)Demirdjian, Hogancamp, and Gunlycke]{demirdjian2025efficientdecompositioncarlemanlinearized}
Reuben Demirdjian, Thomas Hogancamp, and Daniel Gunlycke.
\newblock {Efficient decomposition of the Carleman linearized Burgers' equation}.
\newblock \emph{Phys. Rev. A}, 113:\penalty0 032408, Mar 2026.
\newblock \doi{10.1103/g27q-r2gk}.

\bibitem[Dervovic et~al.(2018)Dervovic, Herbster, Mountney, Severini, Usher, and Wossnig]{dervovic2018quantumlinearsystemsalgorithms}
Danial Dervovic, Mark Herbster, Peter Mountney, Simone Severini, Naïri Usher, and Leonard Wossnig.
\newblock Quantum linear systems algorithms: a primer.
\newblock 2018.
\newblock \doi{10.48550/arXiv.1802.08227}.

\bibitem[Dodin and Startsev(2021)]{Dodin}
I.~Y. Dodin and E.~A. Startsev.
\newblock {On applications of quantum computing to plasma simulations}.
\newblock \emph{Physics of Plasmas}, 28\penalty0 (9):\penalty0 092101, 09 2021.
\newblock ISSN 1070-664X.
\newblock \doi{10.1063/5.0056974}.

\bibitem[Engel et~al.(2019)Engel, Smith, and Parker]{plasmaquant1}
Alexander Engel, Graeme Smith, and Scott~E. Parker.
\newblock Quantum algorithm for the {V}lasov equation.
\newblock \emph{Phys. Rev. A}, 100:\penalty0 062315, Dec 2019.
\newblock \doi{10.1103/PhysRevA.100.062315}.

\bibitem[Engel et~al.(2021)Engel, Smith, and Parker]{embedding}
Alexander Engel, Graeme Smith, and Scott~E. Parker.
\newblock {Linear embedding of nonlinear dynamical systems and prospects for efficient quantum algorithms}.
\newblock \emph{Physics of Plasmas}, 28\penalty0 (6):\penalty0 062305, 06 2021.
\newblock ISSN 1070-664X.
\newblock \doi{10.1063/5.0040313}.

\bibitem[Ferrari(1998)]{ferrari_plasma_jet}
Attilio Ferrari.
\newblock Modeling extragalactic jets.
\newblock \emph{Annual Review of Astronomy and Astrophysics}, 36\penalty0 (Volume 36, 1998):\penalty0 539--598, 1998.
\newblock ISSN 1545-4282.
\newblock \doi{10.1146/annurev.astro.36.1.539}.

\bibitem[Ferri\`ere(2001)]{plasmastuff}
Katia~M. Ferri\`ere.
\newblock The interstellar environment of our galaxy.
\newblock \emph{Rev. Mod. Phys.}, 73:\penalty0 1031--1066, Dec 2001.
\newblock \doi{10.1103/RevModPhys.73.1031}.

\bibitem[Forets and Pouly(2017)]{forets2017explicit}
Marcelo Forets and Amaury Pouly.
\newblock {Explicit Error Bounds for Carleman Linearization}.
\newblock 2017.
\newblock \doi{10.48550/arXiv.1711.02552}.

\bibitem[Gaitan(2020)]{gaitan1}
Frank Gaitan.
\newblock {Finding flows of a Navier--Stokes fluid through quantum computing}.
\newblock \emph{npj Quantum Information}, 6\penalty0 (1):\penalty0 61, 2020.
\newblock \doi{10.1038/s41534-020-00291-0}.

\bibitem[Gaitan(2021)]{gaitan2}
Frank Gaitan.
\newblock {Finding Solutions of the Navier-Stokes Equations through Quantum Computing—Recent Progress, a Generalization, and Next Steps Forward}.
\newblock \emph{Advanced Quantum Technologies}, 4\penalty0 (10):\penalty0 2100055, 2021.
\newblock \doi{10.1002/qute.202100055}.

\bibitem[Giannakis et~al.(2022)Giannakis, Ourmazd, Pfeffer, Schumacher, and Slawinska]{koopman6}
Dimitrios Giannakis, Abbas Ourmazd, Philipp Pfeffer, J\"org Schumacher, and Joanna Slawinska.
\newblock Embedding classical dynamics in a quantum computer.
\newblock \emph{Phys. Rev. A}, 105:\penalty0 052404, May 2022.
\newblock \doi{10.1103/PhysRevA.105.052404}.

\bibitem[Gnanasekaran et~al.(2024)Gnanasekaran, Surana, and Zhu]{variational_Carleman}
Abeynaya Gnanasekaran, Amit Surana, and Hongyu Zhu.
\newblock {Variational Quantum Framework for Nonlinear PDE Constrained Optimization Using Carleman Linearization}.
\newblock 2024.
\newblock \doi{10.48550/arXiv.2410.13688}.

\bibitem[Gonzalez-Conde et~al.(2025)Gonzalez-Conde, Lewis, Bharadwaj, and Sanz]{linearization_efficiency}
Javier Gonzalez-Conde, Dylan Lewis, Sachin~S. Bharadwaj, and Mikel Sanz.
\newblock {Quantum Carleman linearization efficiency in nonlinear fluid dynamics}.
\newblock \emph{Phys. Rev. Res.}, 7:\penalty0 023254, Jun 2025.
\newblock \doi{10.1103/PhysRevResearch.7.023254}.

\bibitem[Griffin et~al.(2019)Griffin, Jain, Flint, and Chan]{navier_stokes}
Kevin Griffin, Suhas Jain, Tim Flint, and Wai Hong~Ronald Chan.
\newblock {Investigation of quantum algorithms for direct numerical simulation of the Navier-Stokes equations}.
\newblock Annual Research Briefs 2019, 12 2019.
\newblock \doi{10.13140/RG.2.2.22657.81762}.

\bibitem[Grover(1996)]{grover}
Lov~K. Grover.
\newblock A fast quantum mechanical algorithm for database search.
\newblock In \emph{Proceedings of the Twenty-Eighth Annual ACM Symposium on Theory of Computing}, STOC '96, page 212–219, New York, NY, USA, 1996. Association for Computing Machinery.
\newblock ISBN 0897917855.
\newblock \doi{10.1145/237814.237866}.

\bibitem[Haack et~al.(2021)Haack, Hauck, Klingenberg, Pirner, and Warnecke]{krook3}
Jeffrey Haack, C.~Hauck, Christian Klingenberg, Marlies Pirner, and Sandra Warnecke.
\newblock A consistent {BGK} model with velocity-dependent collision frequency for gas mixtures.
\newblock \emph{Journal of Statistical Physics}, 184, 09 2021.
\newblock \doi{10.1007/s10955-021-02821-2}.

\bibitem[Harrow et~al.(2009)Harrow, Hassidim, and Lloyd]{harrow}
Aram~W Harrow, Avinatan Hassidim, and Seth Lloyd.
\newblock Quantum algorithm for linear systems of equations.
\newblock \emph{Physical review letters}, 103\penalty0 (15):\penalty0 150502, 2009.
\newblock \doi{10.1103/PhysRevLett.103.150502}.

\bibitem[Horn and Johnson(1985)]{matrix_book}
Roger~A. Horn and Charles~R. Johnson.
\newblock \emph{{Matrix Analysis}}.
\newblock Cambridge University Press, 1985.
\newblock \doi{10.1017/CBO9780511810817}.

\bibitem[Itani and Succi(2022)]{lattice_boltzmann_1}
Wael Itani and Sauro Succi.
\newblock {Analysis of Carleman Linearization of Lattice Boltzmann}.
\newblock \emph{Fluids}, 7\penalty0 (1), 2022.
\newblock ISSN 2311-5521.
\newblock \doi{10.3390/fluids7010024}.

\bibitem[Jennings et~al.(2025)Jennings, Korzekwa, Lostaglio, Sornborger, Subasi, and Wang]{jennings2025quantumalgorithmsgeneralnonlinear}
David Jennings, Kamil Korzekwa, Matteo Lostaglio, Andrew~T Sornborger, Yigit Subasi, and Guoming Wang.
\newblock {Quantum algorithms for general nonlinear dynamics based on the Carleman embedding}.
\newblock 2025.
\newblock \doi{10.48550/arXiv.2509.07155}.

\bibitem[Joseph et~al.(2023)Joseph, Shi, Porter, Castelli, Geyko, Graziani, Libby, and DuBois]{Joseph}
I.~Joseph, Y.~Shi, M.~D. Porter, A.~R. Castelli, V.~I. Geyko, F.~R. Graziani, S.~B. Libby, and J.~L. DuBois.
\newblock {Quantum computing for fusion energy science applications}.
\newblock \emph{Physics of Plasmas}, 30\penalty0 (1):\penalty0 010501, 01 2023.
\newblock ISSN 1070-664X.
\newblock \doi{10.1063/5.0123765}.

\bibitem[Joseph(2020)]{koopman7}
Ilon Joseph.
\newblock Koopman--von {N}eumann approach to quantum simulation of nonlinear classical dynamics.
\newblock \emph{Phys. Rev. Res.}, 2:\penalty0 043102, Oct 2020.
\newblock \doi{10.1103/PhysRevResearch.2.043102}.

\bibitem[Krovi(2023)]{krovi}
Hari Krovi.
\newblock Improved quantum algorithms for linear and nonlinear differential equations.
\newblock \emph{Quantum}, 7:\penalty0 913, 2023.
\newblock \doi{10.22331/q-2023-02-02-913}.

\bibitem[Leyton and Osborne(2008)]{sarah}
Sarah~K Leyton and Tobias~J Osborne.
\newblock A quantum algorithm to solve nonlinear differential equations.
\newblock 2008.
\newblock \doi{10.48550/arXiv.0812.4423}.

\bibitem[Li et~al.(2025)Li, Yin, Wiebe, Chun, Schenter, Cheung, and M\"ulmenst\"adt]{lattice_boltzmann_3}
Xiangyu Li, Xiaolong Yin, Nathan Wiebe, Jaehun Chun, Gregory~K. Schenter, Margaret~S. Cheung, and Johannes M\"ulmenst\"adt.
\newblock {Potential quantum advantage for simulation of fluid dynamics}.
\newblock \emph{Phys. Rev. Res.}, 7:\penalty0 013036, Jan 2025.
\newblock \doi{10.1103/PhysRevResearch.7.013036}.

\bibitem[Liu et~al.(2021)Liu, Øie Kolden, Krovi, Loureiro, Trivisa, and Childs]{liu}
Jin-Peng Liu, Herman Øie Kolden, Hari~K. Krovi, Nuno~F. Loureiro, Konstantina Trivisa, and Andrew~M. Childs.
\newblock Efficient quantum algorithm for dissipative nonlinear differential equations.
\newblock \emph{Proceedings of the National Academy of Sciences}, 118\penalty0 (35):\penalty0 e2026805118, 2021.
\newblock \doi{10.1073/pnas.2026805118}.

\bibitem[Liu et~al.(2023)Liu, An, Fang, Wang, Low, and Jordan]{liu2}
Jin-Peng Liu, Dong An, Di~Fang, Jiasu Wang, Guang~Hao Low, and Stephen Jordan.
\newblock {Efficient Quantum Algorithm for Nonlinear Reaction-Diffusion Equations and Energy Estimation}.
\newblock \emph{Communications in Mathematical Physics}, 404:\penalty0 963--1020, 2023.
\newblock ISSN 1432-0916.
\newblock \doi{10.1007/s00220-023-04857-9}.

\bibitem[Lloyd et~al.(2020)Lloyd, De~Palma, Gokler, Kiani, Liu, Marvian, Tennie, and Palmer]{lloyd}
Seth Lloyd, Giacomo De~Palma, Can Gokler, Bobak Kiani, Zi-Wen Liu, Milad Marvian, Felix Tennie, and Tim Palmer.
\newblock Quantum algorithm for nonlinear differential equations.
\newblock 2020.
\newblock \doi{10.48550/arXiv.2011.06571}.

\bibitem[Miyamoto et~al.(2024)Miyamoto, Yamazaki, Uchida, Fujisawa, and Yoshida]{neutrino}
Koichi Miyamoto, Soichiro Yamazaki, Fumio Uchida, Kotaro Fujisawa, and Naoki Yoshida.
\newblock {Quantum algorithm for the {Vlasov} simulation of the large-scale structure formation with massive neutrinos}.
\newblock \emph{Phys. Rev. Res.}, 6:\penalty0 013200, Feb 2024.
\newblock \doi{10.1103/PhysRevResearch.6.013200}.

\bibitem[Novikau et~al.(2022)Novikau, Startsev, and Dodin]{Novikau}
I.~Novikau, E.~A. Startsev, and I.~Y. Dodin.
\newblock Quantum signal processing for simulating cold plasma waves.
\newblock \emph{Phys. Rev. A}, 105:\penalty0 062444, Jun 2022.
\newblock \doi{10.1103/PhysRevA.105.062444}.

\bibitem[Novikau et~al.(2024)Novikau, Dodin, and Startsev]{novikau2024encoding}
Ivan Novikau, Ilya~Y. Dodin, and Edward~A. Startsev.
\newblock Encoding of linear kinetic plasma problems in quantum circuits via data compression.
\newblock 2024.
\newblock \doi{10.48550/arXiv.2403.11989}.

\bibitem[Parks(1995)]{physics_of_space_plasmas}
George~K Parks.
\newblock \emph{{Physics Of Space Plasmas: An Introduction}}.
\newblock CRC Press, 1 edition, 1995.
\newblock \doi{10.1201/9780429301674}.

\bibitem[Penuel et~al.(2025)Penuel, Katabarwa, Johnson, Kuklinski, Rempfer, Farquhar, Cao, and Garrett]{lattice_boltzmann_5}
John Penuel, Amara Katabarwa, Peter~D. Johnson, Parker Kuklinski, Benjamin Rempfer, Collin Farquhar, Yudong Cao, and Michael~C. Garrett.
\newblock {Detailed assessment of calculating drag force with quantum computers: Explicit time-evolution precludes exponential advantage for nonlinear differential equations}.
\newblock 2025.
\newblock \doi{10.48550/arXiv.2406.06323}.

\bibitem[Press et~al.(2007)Press, Teukolsky, Vetterling, and Flannery]{numerical}
William~H. Press, Saul~A. Teukolsky, William~T. Vetterling, and Brian~P. Flannery.
\newblock \emph{{Numerical Recipes 3rd Edition: The Art of Scientific Computing}}.
\newblock Cambridge University Press, USA, 3 edition, 2007.
\newblock ISBN 0521880688.
\newblock URL \url{https://dl.acm.org/doi/10.5555/1403886}.

\bibitem[Rosenbluth et~al.(1957)Rosenbluth, MacDonald, and Judd]{fokker}
Marshall~N. Rosenbluth, William~M. MacDonald, and David~L. Judd.
\newblock Fokker-{P}lanck equation for an inverse-square force.
\newblock \emph{Phys. Rev.}, 107:\penalty0 1--6, Jul 1957.
\newblock \doi{10.1103/PhysRev.107.1}.

\bibitem[Shor(1997)]{shor}
Peter~W. Shor.
\newblock {Polynomial-Time Algorithms for Prime Factorization and Discrete Logarithms on a Quantum Computer}.
\newblock \emph{SIAM Journal on Computing}, 26\penalty0 (5):\penalty0 1484--1509, 1997.
\newblock \doi{10.1137/S0097539795293172}.

\bibitem[Sircombe and Arber(2009)]{vlasov_solver_3}
N.~J. Sircombe and T.~D. Arber.
\newblock {VALIS}: A split-conservative scheme for the relativistic {2D} {Vlasov}--{Maxwell} system.
\newblock \emph{Journal of Computational Physics}, 228\penalty0 (13):\penalty0 4773--4788, 2009.
\newblock \doi{10.1016/j.jcp.2009.03.029}.

\bibitem[Struchtrup(1997)]{krook2}
Henning Struchtrup.
\newblock The {BGK}-model with velocity-dependent collision frequency.
\newblock \emph{Continuum Mechanics and Thermodynamics}, 9:\penalty0 23--31, 02 1997.
\newblock \doi{10.1007/s001610050053}.

\bibitem[Surana et~al.(2023)Surana, Gnanasekaran, and Sahai]{surana2023efficient}
Amit Surana, Abeynaya Gnanasekaran, and Tuhin Sahai.
\newblock An efficient quantum algorithm for simulating polynomial differential equations.
\newblock 2023.
\newblock \doi{10.48550/arXiv.2212.10775}.

\bibitem[Vaszary(2024{\natexlab{a}})]{arxiv_carleman}
Tamás Vaszary.
\newblock {Carleman Linearization of Partial Differential Equations}.
\newblock 2024{\natexlab{a}}.
\newblock \doi{10.48550/arXiv.2412.00014}.

\bibitem[Vaszary(2024{\natexlab{b}})]{dissertation}
Tamás Vaszary.
\newblock {Solving the Nonlinear Vlasov Equation on a Quantum Computer (Dissertation)}, May 2024{\natexlab{b}}.
\newblock URL \url{https://doi.org/10.5281/zenodo.11200239}.

\bibitem[Wang et~al.(2026)Wang, Jia, Veerapaneni, and Ding]{wang2026quantumalgorithmsnonlineardifferential}
Ke~Wang, Zikang Jia, Shravan Veerapaneni, and Zhiyan Ding.
\newblock {Quantum Algorithms for Nonlinear Differential Equations via Pivot-Shifted Carleman Linearization}.
\newblock 2026.
\newblock \doi{10.48550/arXiv.2605.20071}.

\bibitem[Wossnig et~al.(2018)Wossnig, Zhao, and Prakash]{dense}
Leonard Wossnig, Zhikuan Zhao, and Anupam Prakash.
\newblock {Quantum Linear System Algorithm for Dense Matrices}.
\newblock \emph{Phys. Rev. Lett.}, 120:\penalty0 050502, Jan 2018.
\newblock \doi{10.1103/PhysRevLett.120.050502}.

\bibitem[Wu et~al.(2025)Wu, Wang, and Li]{no_dissipative}
Hsuan-Cheng Wu, Jingyao Wang, and Xiantao Li.
\newblock {Quantum Algorithms for Nonlinear Dynamics: Revisiting Carleman Linearization with No Dissipative Conditions}.
\newblock \emph{SIAM Journal on Scientific Computing}, 47\penalty0 (2):\penalty0 A943--A970, 2025.
\newblock \doi{10.1137/24M1665799}.

\bibitem[Xue et~al.(2021)Xue, Wu, and Guo]{homotopy}
Cheng Xue, Yu-Chun Wu, and Guo-Ping Guo.
\newblock Quantum homotopy perturbation method for nonlinear dissipative ordinary differential equations.
\newblock \emph{New Journal of Physics}, 23\penalty0 (12):\penalty0 123035, dec 2021.
\newblock \doi{10.1088/1367-2630/ac3eff}.

\bibitem[Zylberman et~al.(2022{\natexlab{a}})Zylberman, Di~Molfetta, Brachet, Loureiro, and Debbasch]{Julien}
Julien Zylberman, Giuseppe Di~Molfetta, Marc Brachet, Nuno~F. Loureiro, and Fabrice Debbasch.
\newblock Quantum simulations of hydrodynamics via the {M}adelung transformation.
\newblock \emph{Phys. Rev. A}, 106:\penalty0 032408, Sep 2022{\natexlab{a}}.
\newblock \doi{10.1103/PhysRevA.106.032408}.

\bibitem[Zylberman et~al.(2022{\natexlab{b}})Zylberman, Molfetta, Brachet, Loureiro, and Debbasch]{Julien2}
Julien Zylberman, Giuseppe~Di Molfetta, Marc Brachet, Nuno~F. Loureiro, and Fabrice Debbasch.
\newblock {Hybrid Quantum-Classical Algorithm for Hydrodynamics}.
\newblock 2022{\natexlab{b}}.
\newblock \doi{10.48550/arXiv.2202.00918}.

\bibitem[Óscar Amaro and Cruz(2023)]{living}
Óscar Amaro and Diogo Cruz.
\newblock {A Living Review of Quantum Computing for Plasma Physics}.
\newblock 2023.
\newblock \doi{10.48550/arXiv.2302.00001}.

\end{thebibliography}

\appendix

\section{Mathematical definitions}\label{appendix:maths}
This appendix collects mathematical definitions and identities used throughout the main text. The notation introduced here follows standard conventions in numerical linear algebra and quantum algorithms.

\subsection{Norm definitions}\label{appendix:norms}
The $l_k$, and in particular the $l_2$ norm of vector a $v$ with length $N$ is
\begin{align}
\norm{v}_k \equiv\left( \sum_{n=1}^N |v_n|^k\right)^{1/k},\quad \quad \norm{v} \equiv \norm{v}_2.
\end{align}
Similarly for matrices, the $l_k$ norm, and in particular the $l_2$ spectral norm is defined as
\begin{align}
\norm{A}_k\equiv\max_{\norm{x}_k\neq 0}\frac{\norm{Ax}_k}{\norm{x}_k},\quad \quad \norm{A} \equiv \norm{A}_2.
\end{align}
The $l_1,\: l_2$ and $l_\infty$ norms simplify to
\begin{subequations}
\begin{align}
\norm{A}_1 &\equiv \max_j \sum_i \left| A_{ij}\right |, \label{eq:l1norm}\\
\norm{A} & \equiv \sqrt{\lambda_{\max}\left(A^\dagger A\right)} ,\label{eq:l2norm}\\
\norm{A}_\infty & \equiv\max_i \sum_j \left| A_{ij}\right |,\label{eq:linfnorm}
\end{align}
\end{subequations}
where $\lambda_{\max}(\cdot)$ returns the largest eigenvalue of its input and $\dagger$ denotes the Hermitian conjugate. An alternative measure is the Frobenius norm:
\begin{align}
\norm{A}_F\equiv\sqrt{\sum_{n=1}^N\sum_{m=1}^M|A_{nm}|^2},
\end{align}
for a matrix of size $(N,M)$. The Frobenius norm is useful because it is easy to compute and provides an upper bound on the spectral norm. These matrix norms exist for all matrices. 

The ($l_2$) log-norm of a square matrix $A$ is given by
\begin{align}
\mu\left(A\right)\equiv\lim_{h\to 0^+}\frac{\norm{\mathbb{I}+hA}-1}{h}=\lambda_{\max}\left\{\frac{A+A^\mathsf{T}}{2}\right\}.
\end{align}
The log-norm enters the convergence criteria of the quantum algorithm through the dissipativity condition discussed in Subsection~\ref{sec:convergence_citeria}.

Finally, $\alpha(A)\equiv\max\left\{\text{Re}(\lambda) \right\}$ is the spectral abscissa, the largest one of the real parts of the eigenvalues $\lambda$ of $A$. Unlike $\norm{A}_k$ and $\norm{A}_F$, $\mu\left(A\right)$ and $\alpha\left(A\right)$ can take negative values and are only applicable to square matrices.

The following inequalities are used repeatedly when deriving bounds on matrix norms and convergence parameters.
\begin{subequations}
\begin{align}
\alpha(A)&\leq \mu(A)\leq \norm{A},\label{eq:evaluerelation}\\
\norm{A}&\leq \norm{A}_F, \label{eq:frobenius_inequality}\\
\norm{A}&\leq \sqrt{\norm{A}_1\norm{A}_\infty}.\label{eq:normbound}
\end{align}
 \end{subequations}

\subsection{Vectorization}\label{vectorization}
Vectorization is used throughout the paper to rewrite the finite-difference equations as a finite-dimensional ODE system of the form required by the quantum algorithm. The vectorization operator takes a matrix and outputs a column vector which contains each matrix element once. We choose to work with \emph{row-major} vectorization, that places the elements of the input matrix into the vector row by row going from top to bottom. For an arbitrary matrix $A$ with size $(N,M)$ it acts as
\begin{align}
\text{vec}(A)=\text{vec}
\begin{pmatrix}
A_{1,1} & \cdots &A_{1,M} \\
\vdots & \ddots &\vdots\\
A_{N,1} & \cdots &A_{N,M}\\
\end{pmatrix}
\equiv
\left[
A_{1,1},\:
\dots ,\:
A_{1,M},\:
A_{2,1} ,\:
\dots ,\:
A_{2,M} ,\:
\dots ,\:
A_{N,M}
\right]^\mathsf{T}.\label{eq:vectorizationdef}
\end{align}
The resulting $\text{vec}(A)$ has length $NM$. Note that vectorization is entirely invertible, as long as the dimensions of the target matrix are specified.

\subsection{Direct sum}\label{direct_sum}
Direct sums appear naturally in the construction of the $F^{(2)}$ matrix in the case of coupling Gauss's law with the Vlasov equation, as shown later in Appendix~\ref{F2subsection}. The direct sum takes matrices and places them in a higher dimensional matrix in the form of a block-diagonal chain. For multiple matrices $\left\{A^{(i)}\right\}$ with $i=1,\dots,I$, the direct sum is
\begin{align}
\bigoplus_{i=1}^I A^{(i)} \equiv A^{(1)} \oplus A^{(2)} \oplus \dots \oplus A^{(I)} \equiv
\begin{pmatrix}
A^{(1)}  & 0 & \cdots & 0 \\
0        & A^{(2)} & \cdots & 0 \\
\vdots   & \vdots  & \ddots & \vdots \\
0        & 0       & \cdots & A^{(I)}
\end{pmatrix}.\label{eq:directsumexample2}
\end{align}
The sizes of the zeros can be identified uniquely based on the sizes of the matrices $\left\{A^{(i)}\right\}$. 

\subsection{Identities used}
For completeness, we collect here the matrix identities required in the derivations of Appendices~\ref{appendix:norm_derivations}-\ref{appendix:quantities_for_complexity}.
\begin{enumerate}
    \item\label{identity:1} Transpose of a Kronecker product is the Kronecker product of the transposes: $\left(A\otimes B\right)^\mathsf{T}=A^\mathsf{T}\otimes B^\mathsf{T}$.
    \item\label{identity:2} The eigenvalues of the tensor product $A\otimes B$ are products of eigenvalues $\lambda_i \omega_j$ if $\lambda_i$ is an eigenvalue of $A$ and $\omega_j$ is that of $B$.
    \item\label{identity:3} The only non-zero eigenvalue of the outer product $v v^\mathsf{T}$ of the column vector $v$ with itself is the inner product $\lambda = v^\mathsf{T}v=\norm{v}^2$.
    \item\label{identity:4} The eigenvalues of $F(A)$ are $F(\lambda_i)$ if $F(x)$ has a Taylor expansion near $x=0$.
    \item\label{identity:5} The eigenvalues of the $(n,n)$ sized tridiagonal Toeplitz matrix 
    \begin{align}
    \begin{pmatrix}
    \alpha  &\beta &0&\cdots&0 \\
    \gamma  &\alpha &\beta&\cdots&0\\
    0&\gamma &\alpha &\cdots&\vdots\\
    \vdots&\vdots&\ddots&\ddots&\beta\\
    0&0&\cdots&\gamma&\alpha 
    \end{pmatrix}\label{eq:toeplitzmatrix}
    \end{align}
    are
    \begin{align}
    \lambda_k = \alpha + 2\beta\sqrt{\frac{\gamma}{\beta}}\cos \left(\frac{k\pi}{n+1}\right), \quad \quad k=1,\dots, n.
    \end{align}
\end{enumerate}

\subsection{Discrete calculus}\label{discrete_calculus}
The finite-difference scheme introduced in Subsection~\ref{sec:gauss_law_finite_diff} requires discrete approximations to derivatives and integrals. We summarize here the integration rules used in constructing the matrices $F^{(0)}$ and $F^{(2)}$.

In particular, the double integral in Eq. \eqref{eq:1species_discretized_a}, that is cumulative in $x$ and goes over the whole range of $v$, is approximated with the 2 dimensional trapezoidal rule \cite{numerical}, given by
\begin{equation}
\iint^{x_i}f_{IJ}\:dx_Idv_J\equiv\frac{\Delta x\Delta v}{2}\cdot\\
\begin{cases}
    \begin{cases}
        0,\\
    \end{cases}
    &\begin{aligned}\mkern-0mu\text{for $i=1$,}\end{aligned} \\[7pt]
    \begin{cases}
    \displaystyle\smash[b]{\sum_{J=1}^{N_v}\left(f_{1,J}+f_{i,J}\right)+2\sum_{J=1}^{N_v}\sum_{I=2}^{i-1}f_{I,J}},\\[10pt]
    \end{cases}
    &\begin{aligned}\mkern-0mu\text{for $2\leq i \leq N_x$}.\end{aligned}\label{eq:trapezoidal}
\end{cases}
\end{equation}
Similarly, the single integral in Eq. \eqref{eq:1species_discretized_e_ampere} is approximated as
\begin{align}
\int v_J f_{iJ}\:dv_J \equiv \Delta v \sum_{j=1}^{N_v}v_J f_{iJ}.
\end{align}
Note that despite using the trapezoidal rule, the $j=1$ and $j=N_v$ gridpoints are not treated differently from the others, unlike the $i=1$ and $i=N_x$ ones. This is due to the $v$ integral going over the range $-\infty < v< \infty$, so in theory the $f_{ij}$ values at $|v|>v_{\max}$ velocities (with $f(x,|v|>v_{\max})=0$) also enter the integrals but have vanishing contributions. Hence all of the above approximations carry an error of $\mathcal{O}(\Delta x^2+\Delta v^2)$.

\section{Vectorization of dependent variables}\label{appendix:vectorization_of_u_both}
The purpose of this appendix is to provide the explicit index mappings used to convert the finite-difference equations into the vectorized ODE form required in Eq.~\eqref{eq:krovidef}.

\subsection{Coupling to Gauss's law}\label{appendix:vectorization_of_u_gauss}
In Section \ref{sec:mapping}, we defined the vector $u$ as the row-major vectorization of the distribution matrix $f$. Here we state the conversion formulas, from which elements of $u$ can be expressed from those of $f$ and vice versa. We have
\begin{subequations}
\begin{align}
u_n&= 
f_{\lceil n/N_v\rceil,\: n\sslash N_v}, &&\text{where }\underbrace{\lceil n/N_v\rceil}_{\text{computes row index }i},\quad \underbrace{n\sslash N_v}_{\text{computes column index }j},\quad\label{eq:unfromfij}\\
f_{ij}&=u_{(i-1)N_v+j}, &&\text{where }\underbrace{(i-1)N_v}_{\text{goes over }i-1\text{ full rows}},\quad\underbrace{j}_{\text{gives }j\text{-th element of row }i},\label{eq:fijfromun}
\end{align}
\end{subequations}
where the $\sslash$ notation was defined in Table \ref{tab:notation}. It computes the second index of $f$, from $n$. Note that the ranges of the indices are $i=1,\dots,N_x;\: j=1,\dots,N_v;\: n=1,\dots, N$. The above equations express the one-to-one correspondence between the vector index $n$ and the grid-point indices $(i,j)$. Then, $u^{\otimes2}$ becomes
\begin{align}
u^{\otimes2}=
\left[ f_{1,1}u^\mathsf{T}, \:  \cdots, \: f_{1,N_v}u^\mathsf{T}, \: f_{2,1}u^\mathsf{T}, \: \cdots, \: f_{2,N_v}u^\mathsf{T},\: \dots, \: f_{N_x,N_v}u^\mathsf{T} \right]^\mathsf{T}
,
\end{align}
where each `element' above is an $N$ dimensional vector. For example, the first one reads $f_{11}u=[f_{11}u_1,f_{11}u_2,\dots,f_{11}u_N]^\mathsf{T}$. Hence, $u^{\otimes2}$ is $N^2$ dimensional. Its elements can be associated to those of $u$ and $f$, and vice versa, via
\begin{subequations}
\begin{align}
u^{\otimes 2}_n&=u_{\lceil n/N\rceil}\cdot u_{ n\sslash N}, &&\text{where} \underbrace{\lceil n/N\rceil}_{\text{gives first term of product}},\quad \underbrace{ n\sslash N}_{\text{gives second term of product}},
\label{eq:u2fromuu}\\
u_a  u_b &= u^{\otimes 2}_{N(a-1)+b}, &&\text{where}
\underbrace{N(a-1)}_{\text{goes over terms }u_1 u,\dots u_{a-1}u},\quad\underbrace{b}_{\text{gives }b\text{-th element of }u_a u},
\label{eq:uufromu2}\\
f_{ab} f_{cd}&=u^{\otimes 2}_{N[(a-1)N_v+b-1]+(c-1)N_v+d}, &&\text{where}
\underbrace{N[(a-1)N_v+b-1]}_{\text{goes over terms }f_{1,1}u,\dots f_{a,b-1}u\:\:},\underbrace{(c-1)N_v+d}_{\text{gives element within 
 }f_{a,b}u}.
\label{eq:u2fromff}
\end{align}
\end{subequations}
Note that Eqs. (\ref{eq:uufromu2}-\ref{eq:u2fromff}) are not unique since the Kronecker product places each $f_{ab}\cdot f_{cd}$ term twice into $u^{\otimes2}$, except the square terms. These relations follow directly from the definition of the Kronecker product and the ordering chosen for $u$.

\subsection{Coupling to Ampere's law}\label{appendix:vectorization_of_u_ampere}
In the Ampere's law case the vector $u$ differs from the Gauss-law case only in that the electric-field variables are appended to the distribution-function variables, as shown in Eq.~\eqref{eq:amperestatevector}. In this case $u$ has length $N_xN_v+N_x=N_x(N_v+1)$. We have invertible maps similarly to the Gauss's law case, in the form
\begin{subequations}
\begin{align}
u_n&= \begin{cases}
f_{\lceil n/N_v\rceil,\: n\sslash N_v}, &\text{for}\:\:1\leq n\leq N_x N_v,\\
E_{n- N_x N_v},&\text{else,}
\end{cases}\\
f_{ij}&=u_{(i-1)N_v+j},\label{eq:ampere_map1}\\
E_i&=u_{N_xN_v+i}.
\label{eq:ampere_map2}
\end{align}
\end{subequations}
The ranges of the indices are $i=1,\dots,N_x;\: j=1,\dots,N_v;\: n=1,\dots, N_x(N_v+1)$. Maps involving $u^{\otimes 2}$ are not required for the argument in Section~\ref{sec:ampere}, since the conclusion follows solely from the structure of the linear matrix $F^{(1)}$. To find $F^{(1)}$, we must again follow the procedure described in Section \ref{sec:method_to} and insert the conversion formulas from above into Eqs. \eqref{eq:1species_discretized_b_ampere}, \eqref{eq:1species_discretized_c_ampere} and \eqref{eq:1species_discretized_e_ampere}. We explicitly write the map for the entries of $F^{(1)}$ that arises from this in Appendix \ref{F1_Ampere_subsection}.

\section{Entries of the $F^{(i)}$ matrices}\label{appendix:entries}
This appendix states the explicit matrix-entry maps obtained by applying the procedure outlined in Section~\ref{sec:mapping}.
\subsection{Gauss's law coupling $F^{(0)}$}
From Eq. \eqref{eq:1species_discretized_d}, we realize the elements of $F^{(0)} $ as
\begin{align}
\left[F^{(0)}\right]_n=\frac{\mathcal{N}}{2x_{\max} \Delta v}\Bigg[\sum_{J=1}^{N_v}\exp(-b v_J^2) \Bigg]^{-1}\cdot \nu(v_{n\sslash N_v})\exp(-b v_{n\sslash N_v}^2),\label{eq:F0map}
\end{align}
with $n=1,\dots, N$. The map follows directly from the inhomogeneous term in the finite-difference equation and the vectorization formulas of Appendix~\ref{appendix:vectorization_of_u_gauss}. Note that the $n$ dependence on the RHS only appears in the form of $n\sslash N_v=j$ as expected from a collision operator without $x$ dependence.

\subsection{Gauss's law coupling $F^{(1)}$}\label{F1subsection}
As discussed in Subsection~\ref{section:matrices}, the linear contribution is partitioned into two parts as $F^{(1)}=F^{(1a)}+F^{(1b)}$, where $F^{(1a)}$ encodes the collisional part, Eq. \eqref{eq:1species_discretized_c}, and $F^{(1b)}$ encodes the rest of the linear contributions, Eq. \eqref{eq:1species_discretized_b}. Therefore,
\begin{align}
\left[F^{(1a)}\right]_{n, k} = -\nu(v_{n\sslash N_v})\delta_{k,n},\label{eq:F1bar}
\end{align}
where the $\delta_{k,n}$ encodes the proportionality with $f_{ij}$. Hence $ F^{(1a)}$ is diagonal. Note that $n,k=1,\dots,N$. For $F^{(1b)}$, the elements read
\begin{align}
\begin{split}
\left[F^{(1b)}\right]_{n,k}=&-\frac{v_{n \sslash N_v}}{2\Delta x}
\begin{cases}
    \delta_{k,n+N_v}-\delta_{k,n+N_v(N_x-1)},&\text{for } n \leq N_v,\\
    \delta_{k,n+N_v}-\delta_{k,n-N_v} , &\text{for } N_v<n \leq N_v\left(N_x-1\right),\\
    \delta_{k,n-N_v(N_x-1)}-\delta_{k,n-N_v}, &\text{else, }
\end{cases}\\
&+\frac{q^2\mathcal{N}}{2m_e\varepsilon_0 \Delta v}\frac{\lceil n/N_v\rceil-1}{N_x}
\begin{cases}
    \delta_{k,n+1}, & \text{for } n \sslash N_v=1 ,\\
    -\delta_{k,n-1},& \text{for } n \sslash N_v=N_v,\\
    \delta_{k,n+1}-\delta_{k,n-1}, & \text{else}.\label{eq:F1hatmap}
\end{cases}
\end{split}
\end{align}
The first term encodes the discretized advection operator $-v_j\,\partial f/\partial x$, while the second term encodes the linear contribution arising from the background charge density. The different cases correspond to the boundary and interior points of the finite-difference stencils.

\subsection{Gauss's law coupling $F^{(2)}$}\label{F2subsection}

The building unit of the $F^{(2)}$ matrix is the trapezoidal rule row vector, denoted by $\mathbb{T}$, that contains all the information about the numerical integration scheme. Just as the integration scheme in Eq. \eqref{eq:trapezoidal}, $\mathbb{T}^{[i]}$ carries a free index $i=1,\dots,N_x$, which is the last position index still being integrated over. Mathematically, it reads
\begin{equation}
\mathbb{T}^{[i]}\equiv\\
\begin{cases}
    \begin{cases}
        \mathsf{T}\:\text{vec}(0_{N_x,N_v})=[\underbrace{0,0\dots ,0}_{N_xN_v\text{ times}}],\\
    \end{cases}
    &\begin{aligned}\text{for $i=1$},\end{aligned} \\[13pt]
    \begin{cases}
    \mathsf{T}\:\text{vec}
\begin{bmatrix}
2 & 2&\cdots&2&2 \\
4 & 4&\cdots&4&4\\
\vdots&\vdots&\ddots&\vdots&\vdots\\
4&4&\cdots&4&4\\
2 & 2&\cdots&2&2\\
0&0&\cdots&0&0\\
\vdots&\vdots&\ddots&\vdots&\vdots\\
0&0&\cdots&0&0
\end{bmatrix}
\begin{matrix}
    \coolrightbrace{1 \\ 2\\ \vdots\\2\\1 }{\phantom{N_x-}\:\,i\text{ rows}}\\
    \coolrightbrace{0 \\ \vdots \\ 0 }{N_x-i\text{ rows}}\\
\end{matrix}\\[68pt]
=[\underbrace{2,\dots ,2}_{N_v\text{ times}},\underbrace{4,\dots ,4}_{N_v(i-2)\text{ times}},\underbrace{2,\dots ,2}_{N_v\text{ times}},\underbrace{0,\dots ,0}_{(N_x-i)N_v\text{ times}}],
    \end{cases}
    &\begin{aligned}\text{for $i>1.$}\end{aligned}\\[14pt]
\end{cases}\label{eq:Tidefinition}
\end{equation}
Here the transpose $\mathsf{T}$ was put in front of the vectorization for readability. For $i=1$, all slots are 0, hence $\mathbb{T}^{[1]}$ is just a vector of 0's. For all $i$, $\mathbb{T}^{[i]}$ is row vector of length $N$. The point of this object is that the electric field generated by the electrons at position $x_i$ can be computed by matrix multiplying $\mathbb{T}^{[i]}$ by $u$ as
\begin{align}
E(x_i)^{(\text{e})}\propto \iint^{x_i}f_{IJ}\:dx_Idv_J=\frac{\Delta x\Delta v}{4}\: \mathbb{T}^{[i]}u.\label{eq:matrixmultipweighting}
\end{align}
The above equation essentially says, that one can force element-wise array multiplication of two matrices (the trapezoidal rule and $f$ in this case) to be a matrix multiplication by vectorizing both (into $\mathbb{T}$ and $u$ in this case).

Using this one may express $F^{(2)}$ as a block-diagonal sequence as
\begin{align}
F^{(2)}=-\frac{q^2\Delta x}{8m_e\varepsilon_0}\cdot\bigoplus_{i=1}^{N_x} B^{[i]} =-\frac{q^2\Delta x}{8m_e\varepsilon_0} \cdot B^{[1]}\oplus B^{[2]}\oplus\dots\oplus B^{[N_x]},\label{eq:onespeciessequence}
\end{align}
where the direct sum is defined in Appendix \ref{direct_sum}. The block $B^{[i]}$ is given by
\begin{align}
B^{[i]}
\equiv
\begin{pmatrix}
0 & 1 & & &\\
-1 & 0&1 & &\\
&\ddots&\ddots &\ddots &\\
&&-1&0&1\\
& & & -1& 0
\end{pmatrix}\otimes \mathbb{T}^{[i]}
,\label{eq:onespeciesBi}
\end{align}
where the left term in the tensor product has size $(N_v,N_v)$. The upper and lower diagonals encode the two `legs' of the finite difference derivative stencil of $\partial f/\partial v$. Together with $ \mathbb{T}^{[i]}$ encoding the double integral of $f$, $B^{[i]}$ compactly expresses the nonlinearity in Eq. \eqref{eq:1species_discretized_a}.

The explicit expression for the entries of $F^{(2)}$ is lengthy because it simultaneously encodes the finite-difference derivative in velocity and the cumulative trapezoidal-rule integration appearing in Gauss's law. The cases are explained as follows. Recall that the state vector index $n$ and the grid indices $(i,j)$ are related by $\lceil n/N_v\rceil=i$, $n\sslash N_v=j$ and $n=(i-1)N_v+j$. The row and column indices are in the ranges $n=1,\dots,N$ and $k=1,\dots,N^2$, respectively.

\begin{enumerate}
    \item Case $\{n< N_v+1\}$ evolves $f_{ij}$ with $i=1$, when the integral of the ion background is 0 by definition.
    \item Case $\left\{\left\lceil n/N_v\right\rceil\geq2\:\: \text{and}\:\:n\sslash N_v= 1\right\}$  evolves $f_{ij}$ with $i\geq 2$ and $j=1$, where the left `leg' of $\partial/\partial v|_{ij}f$ vanishes due to BCs.
    \item Case $\left\{\left\lceil n/N_v \right\rceil\geq2\:\: \text{and}\:\:n\sslash N_v= N_v\right\}$ evolves $f_{ij}$ with $i\geq 2$ and $j=N_v$, where the right `leg' of $\partial/\partial v|_{ij}f$ vanishes due to BCs.
    \item The last case is the general one, in which both `legs' exist.
\end{enumerate}
 
\begin{equation}
\left[F^{(2)}\right]_{n,k}=\shortminus\frac{q^2\Delta x}{4m_e\varepsilon_0}\cdot
\begin{cases}
    \phantom{\shortminus}\begin{cases}
        0,\\
    \end{cases}
    &\begin{aligned}\text{for $n< N_v+1$,}\end{aligned} \\[7pt]
    \phantom{\shortminus}\begin{cases}
    \displaystyle\smash[b]{\sum_{J=1}^{N_v}\left(\delta_{k,Nn+J}+\delta_{k,Nn+(\lceil n/N_v\rceil-1)N_v+J}\right)}\\[12pt]
    \displaystyle\smash[b]{+2\sum_{I=2}^{\lceil n/N_v\rceil-1}\sum_{J=1}^{N_v}\delta_{k,Nn+(I-1)N_v+J}},\\[10pt]
    \end{cases}
    &\begin{aligned} \text{for $\left\lceil\frac{n}{N_v}\right\rceil\geq2$\:\:\:}\\
     \text{and $n\sslash N_v= 1$,}\end{aligned}\\[35pt]
    \shortminus\begin{cases} 
    \displaystyle\smash[b]{\sum_{J=1}^{N_v}\left(\delta_{k,N(n-2)+(\lceil n/N_v\rceil-1)N_v+J}+\delta_{k,N(n-2)+J}\right)}\\[10pt]
    \displaystyle\smash[b]{+2\sum_{I=2}^{\lceil n/N_v\rceil-1}\sum_{J=1}^{N_v}\delta_{k,N(n-2)+(I-1)N_v+J}},\\[10pt]
    \end{cases}
    &\begin{aligned}\text{for $\left\lceil\frac{n}{N_v}\right\rceil\geq2$\:\:\:\:\:\:}\\
    \text{and $n\sslash N_v= N_v$,}\end{aligned}\\[35pt]
    \phantom{\shortminus}\begin{cases}
    \displaystyle\smash[b]{\Bigg[\sum_{J=1}^{N_v}\left(\delta_{k,Nn+J}+\delta_{k,Nn+(\lceil n/N_v\rceil-1)N_v+J}\right)}\\[0pt] \\
    \displaystyle\smash[b]{+2\sum_{I=2}^{\lceil n/N_v\rceil-1}\sum_{J=1}^{N_v}\delta_{k,Nn+(I-1)N_v+J}\Bigg]-}\\[12pt]
    \Bigg[ \displaystyle\smash[b]{\sum_{J=1}^{N_v}\left(\delta_{k,N(n-2)+(\lceil n/N_v\rceil-1)N_v+J}+\delta_{k,N(n-2)+J}\right)} \\[14pt]
    \displaystyle\smash[b]{+2\sum_{I=2}^{\lceil n/N_v\rceil-1}\sum_{J=1}^{N_v}\delta_{k,N(n-2)+(I-1)N_v+J}\Bigg]},\\[5pt]
    \end{cases}
    &\begin{aligned}\text{else,}\end{aligned}
\end{cases}
\end{equation}

\subsection{Ampere's law coupling $F^{(1)}$}\label{F1_Ampere_subsection}
Again, the collision part, Eq. \eqref{eq:1species_discretized_c_ampere}, is packaged into $F^{(1a)}$ and the rest of the linear evolution, Eqs. (\ref{eq:1species_discretized_b_ampere},\ref{eq:1species_discretized_e_ampere}), is packaged into $F^{(1b)}$. Hence $F^{(1)}=F^{(1a)}+F^{(1b)}$ as in the Gauss's law case and we have
\begin{equation}
    \left[F^{(1a)}\right]_{n, k}= 
    \begin{cases}
-\nu(v_{n\sslash N_v})\delta_{k,n}, \hspace{0.5cm} &\text{for } n \leq N_vN_x, \\
0 , &\text{else,} 
\end{cases}
\end{equation}
and
\begin{align}
\left[F^{(1b)}\right]_{n,k}=\begin{cases}
    -\dfrac{v_{n \sslash N_v}}{2\Delta x}\cdot
    \begin{cases}
    \delta_{k,n+N_v}-\delta_{k,n+N_v(N_x-1)}, \hspace{1.1 cm} & \text{for } n \leq N_v, \\
    \delta_{k,n+N_v}-\delta_{k,n-N_v},  & \text{for } N_v<n \leq N_v\left(N_x-1\right),\\
    \delta_{k,n-N_v(N_x-1)}-\delta_{k,n-N_v}, & \text{for } N_v\left(N_x-1\right)<n \leq N_v N_x,
    \end{cases}\\[15 pt]
    \displaystyle\smash[b]{\dfrac{\Delta v q}{\varepsilon_0}\sum_{J=1}^{N_v} \delta_{k,(n-N_xN_v-1)N_v+J}\:v_J}, \hspace{2.35 cm}\text{else,}\\[10 pt]
\end{cases}
\label{F1_map_ampere}
\end{align}
with $n,k=1,\dots N_x(N_v+1)$. The first three cases encode the discretized spatial derivative, while the last line represents the velocity integral appearing in Ampere's law.

\section{Derivation of norms from Gauss's law coupling}\label{appendix:norm_derivations}
This appendix derives the quantities entering the convergence criteria and complexity bounds of the quantum algorithm when coupled to Gauss's law.

\subsection{Bounding $\|F^{(0)}\|$}\label{section:F0_norm}
The vector $F^{(0)}$ originates from the Maxwellian source term and only its asymptotic scaling is required in the main text. Below, we focus on deriving its scaling with $N_x$ and $N_v$. Making use of the fact that $N_v$ is even and writing the $l_2$ norm of Eq. \eqref{eq:F0map} in terms of $j$ only, we get
\begin{align}
\norm{F^{(0)}}
=\frac{\mathcal{N}}{2x_{\max} \Delta v} \Bigg[2\sum_{J=N_v/2+1}^{N_v}\exp(-b v_J^2) \Bigg]^{-1}
\sqrt{2N_x\sum_{j=N_v/2+1}^{N_v} \nu^2(v_j)\exp(-2b v_j^2)}.
\label{eq:F0normexactly}
\end{align}
We bound this part by part: first the $J$ sum from below, or equivalently its reciprocal from above as
\begin{align}
\begin{split}
\sum_{J=N_v/2+1}^{N_v}\exp(-b v_J^2)
&\geq \int_{N_v/2+1}^{N_v+1}\exp(-b v_J^2)dJ\\
&= \frac{\sqrt{\pi}}{2\Delta v\sqrt{b}}\left[\text{erf}\left(\Delta v\sqrt{b}\dfrac{N_v+1}{2}\right)-\text{erf}\left(\Delta v\sqrt{b}\dfrac{1}{2}\right)\right].
\label{eq:boundingMc}
\end{split}
\end{align}
The sum over $j$ must be upper bounded. Recall that $\mathcal{O}(\nu [v_{\text{max}}])=\mathcal{O}(\nu_0)$ due to Eq. \eqref{eq:h_small}, so that we can pull $\nu(v_j)$ out of the sum to get the scaling:
\begin{align}
\begin{split}
\sum_{j=N_v/2+1}^{N_v} \exp(-2b v_j^2)&\leq  \int_{N_v/2}^{N_v}\exp(-2bv_j^2) dj\\
& = \frac{1}{2\sqrt{b}\Delta v}\frac{\sqrt{2\pi}}{2}\left[\text{erf}\left(\Delta v\sqrt{b/2}(N_v-1)\right)+\text{erf}\left(\Delta v\sqrt{b/2}\right)\right].\label{eq:F0_norm_2}
\end{split}
\end{align}
Putting Eq. \eqref{eq:boundingMc} and Eq. \eqref{eq:F0_norm_2} together, we get
\begin{align}
\begin{split}
\lim_{N_x,\:N_v\gg 1}\norm{F^{(0)}} &= \frac{\mathcal{N}N_v \nu_0}{4x_{\max} v_{\max}} \left[2 \frac{N_v\sqrt{\pi}}{4 v_{\max}\sqrt{b}}\text{erf}\left( v_{\max}\sqrt{b}\right)\right]^{-1} \sqrt{2N_x \frac{\sqrt{2\pi}N_v}{8\sqrt{b}v_{\max}}\text{erf}\left(v_{\max}\sqrt{2b}\right)}\\
& = \frac{\mathcal{N}}{x_{\max}\sqrt{v_{\max}}}\left(\frac{b}{2^7 \pi}\right)^{1/4}\frac{\sqrt{\text{erf}\left(v_{\max}\sqrt{2b}\right)}}{\text{erf}\left(v_{\max}\sqrt{b}\right)}\nu_0 \sqrt{N_xN_v}\\
&=\mathcal{O}\left(\nu_0 \sqrt{N_xN_v}\right).
\end{split}
\end{align}

\subsection{Bounding $\|F^{(1)}\|$}\label{section:F1_norm}
We require only an upper bound on $\|F^{(1)}\|$, which is sufficient for the complexity analysis. We use the $l_1$ and $l_\infty$ norms from Eqs. (\ref{eq:l1norm}-\ref{eq:linfnorm}) and the fact that the offdiagonal part of $F^{(1)}$ is antisymmetric, so $F^{(1)}_{ij}=(-1)^{\delta_{ij}+1}F^{(1)}_{ji}$. Then we apply this to the identity Eq. \eqref{eq:normbound} and get
\begin{align}
\begin{split}
\norm{F^{(1)}}&\leq \sqrt{\norm{F^{(1)}}_1 \norm{F^{(1)}}_\infty}\\
& = \norm{F^{(1)}}_1\\
&\leq \nu(v_{\max}) + \frac{q^2\mathcal{N}}{2m_e\varepsilon_0\Delta v}\frac{N_x-1}{N_x}\cdot 2 + \frac{v_{\max}}{2\Delta x}\cdot 2\\
&\leq \mathcal{O}(N_v^{3/2}+N_x),
\end{split}
\end{align}
where $\nu_0\geq \mathcal{O}(N_v^{3/2})$ was substituted from Eq. \eqref{eq:nu_0_scaling} to keep $R\lesssim 1$.

\subsection{Computing $\|F^{(2)}\|$}\label{appendix:F2_norm}
The norm of $F^{(2)}$ controls the strength of the nonlinearity in the convergence parameter $R$ and also affects the complexity estimates. We first write the spectral norm of Eq. \eqref{eq:onespeciessequence} according to its definition:
\begin{align}
\begin{split}
\norm{F^{(2)}}^2&=\left(\frac{q^2\Delta x}{8m_e\varepsilon_0}\right)^2 \lambda_{\max}\left\{ 
\bigoplus_{i=1}^{N_x} {B^{[i]}}^\mathsf{T} B^{[i]}
\right\}\\
&=\left(\frac{q^2\Delta x}{8m_e\varepsilon_0}\right)^2 \max_{i=1,\dots,N_x} 
\norm{B^{[i]}}^2,\label{eq:F2_stuff1}
\end{split}
\end{align}
and then
\begin{align}
\begin{split} 
\norm{B^{[i]}}^2 &= \lambda_{\max}\left\{ 
 {B^{[i]}}^\mathsf{T} B^{[i]}
\right\}\\
 &=-\lambda_{\min}\left\{
\begin{pmatrix}
0 & 1 & & &\\
-1 & 0&1 & &\\
&\ddots&\ddots &\ddots &\\
&&-1&0&1\\
& & & -1& 0
\end{pmatrix}^2
\right\}
\cdot 
\norm{\mathbb{T}^{[i]}}^2,\label{eq:F2_stuff_2}
\end{split}
\end{align}
where Identities \hyperref[identity:1]{1}-\hyperref[identity:3]{3} were applied successively to separate the tensor-product factors. Now from Eq. \eqref{eq:Tidefinition} we can obtain 
\begin{align}
\norm{\mathbb{T}^{[i]}}^2 &= 
\begin{cases}
    0, & \text{for } i=1, \\
    2N_v\cdot 2^2+N_v(i-2)\cdot 4^2, & \text{else. }\\
\end{cases}\label{eq:F2_stuff_4}
\end{align}
Using Identities \hyperref[identity:4]{4} and \hyperref[identity:5]{5}, we also get
\begin{align}
\begin{split}
\lambda_{\min}\left\{
\begin{pmatrix}
0 & 1 & & &\\
-1 & 0&1 & &\\
&\ddots&\ddots &\ddots &\\
&&-1&0&1\\
& & & -1& 0
\end{pmatrix}^2
\right\}&=\min_{k=1,\dots,N_v}\left\{\left(0+2\sqrt{-1}\cos\left(\frac{k\pi}{1+N_v}\right)
\right)^2\right\}\\
&=-4\cos^2 \left(\frac{\pi}{1+N_v}\right).\label{eq:F2_stuff3}
\end{split}
\end{align}
Therefore substituting Eq. \eqref{eq:F2_stuff_4} and Eq. \eqref{eq:F2_stuff3} into Eq. \eqref{eq:F2_stuff_2} results in
\begin{align}
\norm{B^{[i]}}^2 = \begin{cases}
    0, & \text{for } i=1, \\
    32 \cos^2 \left(\dfrac{\pi}{1+N_v}\right) N_v(2i-3) ,& \text{else. }\\
\end{cases}
\end{align}
Putting this back into Eq. \eqref{eq:F2_stuff1} and taking the square root we arrive at
\begin{align}
\norm{F^{(2)}}
&=\frac{q^2 x_{\max}}{\sqrt{2}m_e\varepsilon_0} 
 \cos \left(\frac{\pi}{1+N_v}\right) \frac{\sqrt{N_v(2N_x-3)}}{N_x},
\end{align}
which asymptotically approaches
\begin{align}
\lim_{N_x,\:N_v \gg 1}\norm{F^{(2)}} =
\frac{q^2 x_{\max}}{m_e\varepsilon_0}\sqrt{\frac{N_v}{N_x}}=\mathcal{O}\left(\sqrt{\frac{N_v}{N_x}}\right).
\end{align}

For reference, the Frobenius norm, that upper bounds the spectral norm (Eq. \eqref{eq:frobenius_inequality}) and significantly more straightforward to compute, scales asymptotically as
\begin{align}
\lim_{N_x,\:N_v\gg 1}\norm{F^{(2)}}_{F}=\mathcal{O}(N_v),
\end{align}
which is polynomially larger than the spectral norm and would put even more severe restrictions on the convergence of the quantum algorithm than the one found in Subsection \ref{sec:connection_to_plasma}.

\subsection{Computing $\norm{u_{\text{in}}}$}\label{section:u0_norm}
The norm of the initial condition enters both the convergence parameter and the rescaling procedure discussed in Subsection~\ref{rescaling}. The map corresponding to Eq. \eqref{eq:f(0)_matrix} is
\begin{align}
\Big[u_{\text{in}}\Big]_n = \frac{\mathcal{N}}{2 x_{\max}\Delta v}\left(
\delta_{n\sslash N_v,J}+\delta_{n\sslash N_v,N_v-J+1}\right),
\end{align}
with an associated norm
\begin{align}
\norm{u_{\text{in}}}=
\frac{\mathcal{N}\sqrt{N_x}}{\sqrt{2}x_{\max}\Delta v}.\label{eq:unorm}
\end{align}
Asymptotically, it scales as
\begin{align}
\lim_{N_x,\:N_v\gg 1}\norm{u_{\text{in}}}=
\frac{\mathcal{N}\sqrt{N_x}N_v}{2\sqrt{2}v_{\max}x_{\max}}=\mathcal{O}\left(\sqrt{N_x}N_v\right).\label{eq:u0normscale}
\end{align}

\section{Derivation of quantities entering the complexity analysis}\label{appendix:quantities_for_complexity}
This appendix derives the asymptotic quantities that enter the complexity bounds presented in Section~ \ref{gatecomplexity}.

\subsection{Scaling with $\gamma$}
The rescaling parameter $\gamma$ fixes the normalization of the nonlinear ODE system and therefore enters several quantities used in the complexity analysis. We now evaluate its asymptotic behaviour in the large-grid limit. Substituting the results of the preceding subsections into Eq.~\eqref{eq:gamma} gives
\begin{align}
\lim_{N_x,\:N_v\gg 1} \gamma^2 &=  \frac{\mathcal{N}m_e\varepsilon_0 N_x \sqrt{N_v}}{4\sqrt{2}v_{\max} x_{\max}^2q^2}\left(
\nu_0+\sqrt{\nu_0^2-\left(\frac{b}{2^7 \pi}\right)^{1/4}\frac{\sqrt{\text{erf}\left(v_{\max}\sqrt{2b}\right)}}{\text{erf}\left(v_{\max}\sqrt{b}\right)}\frac{4q^2\mathcal{N}}{m_e\varepsilon_0\sqrt{v_{\max}}}\nu_0 N_v}
\right).
\end{align}
If we again substitute $\nu_0\geq \mathcal{O}(N_v^{3/2})$, the expression simplifies heavily and the second term in the square root vanishes, giving
\begin{align}
\lim_{N_x,\:N_v\gg 1} \gamma = \frac{\mathcal{N}}{2\sqrt{2}v_{\max} x_{\max}}N_x^{1/2} N_v = \mathcal{O}(N_x^{1/2} N_v).
\end{align}
We see that $\gamma$ approaches $\norm{u_{\text{in}}}$ asymptotically.

\subsection{Solution decay}
The fraction $\norm{u_{\text{in}}}/\norm{u(T)}$, denoted by $g_u$, measures the decay of the solution norm. Since it is invariant under the rescaling, we may ignore the latter in the present calculation. To estimate this, we take $u(T)$ to be a Maxwellian, which is a very good approximation for $T\gtrsim 1/\nu_0$. Remember that $F^{(0)}$ in Eq. \eqref{eq:F0map} encodes a Maxwellian that is weighted by values of $\nu(v)$. Therefore, if we set $\nu_0=1$ in $\Vert F^{(0)}\Vert$, the norm, we end up with $\norm{u(T)}=\mathcal{O}(N_v^{1/2}N_x^{1/2})$. Then, substituting $\norm{u_{\text{in}}}$ from Eq. \eqref{eq:u0normscale} results in
\begin{align}
g_u=\frac{\norm{u_{\text{in}}}}{\norm{u(T)}}\approx\frac{\mathcal{O}\left(N_v N_x^{1/2}\right)}{\mathcal{O}\left(N_v^{1/2}N_x^{1/2}\right)}=\mathcal{O}\left(N_v^{1/2}\right).
\end{align}

\subsection{Carleman steps}\label{section:Carleman_steps}
The number of Carleman linearization steps $N_C(\delta)$ is defined in Eq. \eqref{eq:parameters_chosen}. The leading-order contribution produces $\log(1)$ in the denominator. To resolve this, we retain the next-order terms in the relevant asymptotic expansions, such as
\begin{align}
\begin{split}
\nu_0 &\geq \mathcal{O}\left(N_v^{3/2}+N_v\right),\\
-\mu\left(F^{(1)}\right) &= \nu_0+\mathcal{O}(1)=\mathcal{O}\left(N_v^{3/2}+N_v\right),\\
\norm{u_{\text{in}}} & =\mathcal{O}\left(N_x^{1/2}N_v\right),\\
\left(\frac{1}{\norm{\Bar{u}_{\text{in}}}}\right)^2 &= \frac{-\mu\left(F^{(1)}\right)+\sqrt{\mu\left(F^{(1)}\right)^2-4\norm{F^{(2)}}\norm{F^{(0)}}}}{2\norm{u_{\text{in}}}\norm{F^{(2)}}}\\
&=\mathcal{O}\left(1+N_x^{-1}+N_v^{-1/3}\right),\\
\log(1/\norm{\Bar{u}_{\text{in}}}) &= \mathcal{O}\left(N_x^{-1}+N_v^{-1/3}\right).
\end{split}
\end{align}
The second-order scalings of $\Vert F^{(2)}\Vert$ and $\Vert F^{(0)}\Vert$ are not required explicitly, as their leading-order contributions already determine the next-order behaviour of the full expression. Note that the second order contributions originate from terms such as $1/\Delta v=2v_{\max}/(N_v-1)=\mathcal{O}(N_v^{-1}+N_v^{-2})$. Substituting the above scalings into Eq. \eqref{eq:parameters_chosen} yields
\begin{align}
N_C(\delta) =\mathcal{O}\left(\left(N_x+N_v^{1/3}\right)\log\left(\frac{TN_v^2}{\delta}\right)\right).\label{eq:N_C}
\end{align}

\subsection{Norm of the Carleman matrix}
The norm of the Carleman matrix appears explicitly in the QLSA complexity bounds and therefore must be estimated asymptotically. Recall that Eq. (\ref{eq:Anorm}) bounds $\norm{A}$. To compute that, we first perform the rescaling discussed in Subsection \ref{rescaling}:
\begin{align}
\begin{split}
\norm{\Bar{F}^{(2)}} &= \mathcal{O}\left( N_v^{1/2}N_x^{-1/2} N_x^{1/2}N_v\right) = \mathcal{O}\left( N_v^{3/2}\right),\\
\norm{\Bar{F}^{(0)} }&= \mathcal{O}\left( \frac{\nu_0 N_x^{1/2}N_v^{1/2}}{N_x^{1/2}N_v}\right) = \mathcal{O}\left( N_v \right),
\end{split}
\end{align}
where we used $\nu_0\geq \mathcal{O}(N_v^{3/2})$. Then we obtain
\begin{align}
\begin{split}
\norm{A}&\leq N_c(\delta) \Big(
\underbrace{\mathcal{O}\left(N_v\right)}_{\norm{\Bar{F}^{(0)}}}\:\:
+\:\:\underbrace{\mathcal{O}\left(N_v^{3/2}+N_x\right)}_{\norm{F^{(1)}}}\:\:
+\:\:\underbrace{\mathcal{O}\left(N_v^{3/2}\right)}_{\norm{\Bar{F}^{(2)}}}
\Big)\\
&=\mathcal{O}\left(\left(N_x^2+N_v^{11/6}+N_xN_v^{3/2}\right)\log\left(\frac{TN_v^2}{\delta}\right)
\right),
\end{split}
\end{align}
where $\Vert F^{(1)}\Vert$ is unaffected by the rescaling and provides the dominant contribution, and where $N_C(\delta)$ from Eq.~\eqref{eq:N_C} has been substituted.

\subsection{Sparsity of the $F^{(k)}$ matrices}\label{sec:sparsities}
The sparsity of $\Bar{F}^{(0)}$ is $N$ since all its entries are generally nonzero, that of $F^{(1)}$ is $5$ and that of $\Bar{F}^{(2)}$ is $2N$ due to the 2 double integrals encoded per row. Hence,
\begin{align}
s=\max \Big\{N,5,2N \Big\}=2N=\mathcal{O}(N).
\end{align}
The sparsity of the Carleman matrix is inherited from the primitive matrices and is given by
\begin{align}
s_A=3 N_C(\delta) s.
\end{align}

\subsection{Estimating the $\Omega(\delta',\delta)$ factor}
The expression $\Omega(\delta',\delta)$ is defined in Eq. \eqref{eq:parameters_chosen}. In the asymptotic limit it takes the value
\begin{align}
\Omega(\delta',\delta)=\mathcal{O}\left(
\frac{T^2N_v^{3/2}}{\delta'}\left(N_x^2+N_v^{11/6}+N_xN_v^{3/2}\right)\log\left(\frac{TN_v^2}{\delta}\right)
\right).
\end{align}

\subsection{Complexity in terms of grid size}\label{sec:compl_in_terms_of_Nx_Nv_N}
We now substitute the asymptotic estimates derived above into the generic complexity bounds of the quantum algorithm, namely Eq. (\ref{eq:querycomplexity}-\ref{eq:gatecomplexity_factor}), and simplify the resulting expressions. We use the identity $\log(X^aY^b)=\mathcal{O}(\log(XY))$ as $X,Y\to\infty$ with fixed powers $a,b>0$  to simplify our expressions. This gives us the query complexity
\begin{align}
\begin{split}
\mathcal{O}\Bigg(TN_xN_v^{3/2}&\left(
N_x^{7/2}+N_x^{3/2}N_v^{11/6}+N_x^{5/2}N_v^{3/2}+N_x^2N_v^2+N_xN_v^{7/2}+N_v^{23/6}
\right)\times \\
\times&\log^{\frac{3}{2}}\left(\frac{TN_v}{\delta}\right)
 \frac{\log \Omega(\delta',\delta)}{\log \log \Omega(\delta',\delta)}
\text{poly}\left[\left(N_x+N_v^{1/3}\right)\log\left(\frac{TN_v}{\delta}\right)\right]
\times\\
\times&\text{polylog}\Bigg[
\log\Omega(\delta',\delta),T \left(N_x+N_v\right)\log\left(\frac{TN_v}{\delta}\right),N_xN_v,\frac{1}{\varepsilon_q}
\Bigg]
\Bigg),
\label{eq:N_xN_v_querycomplexity}
\end{split}
\end{align}
and the gate complexity that is larger by a factor of
\begin{align}
\mathcal{O}\left(
\text{polylog}\left[\log \Omega(\delta',\delta),
T \left(N_x+N_v\right)\log\left(\frac{TN_v}{\delta}\right),
\frac{1}{\varepsilon_q}
\right]
\right),
\end{align}
where asymptotically we have
\begin{align}
\log \Omega(\delta',\delta)=\mathcal{O}\left(
\log\left[
\frac{TN_v}{\delta'}(N_x+N_v)
\log\left(\frac{TN_v}{\delta}\right)
\right]
\right).
\end{align}
Now, for a grid with a fixed $N_x$ and $N_v$ ratio, i.e substituting $N_x^2=\mathcal{O}(N)=N_v^2$, the query complexity simplifies to
\begin{align}
\begin{split}
\mathcal{O}\Bigg(&TN^{7/2} \log^{\frac{3}{2}}\left(\frac{TN}{\delta}\right)\frac{\log \Omega(\delta',\delta)}{\log \log \Omega(\delta',\delta)}\times \\
&\times\text{poly}\Bigg[
N^{1/2}\log\left(\frac{TN}{\delta}\right)\Bigg]
\text{polylog}\Bigg[
\log \Omega(\delta',\delta),
TN\log\left(\frac{TN}{\delta}\right),N,\frac{1}{\varepsilon_q}
\Bigg]\Bigg),
\label{eq:N_xN_v_querycomplexity_simplified_square}
\end{split}
\end{align}
and the gate complexity to the above times the factor
\begin{align}
\mathcal{O}\left(
\text{polylog}\Bigg[
\log \Omega(\delta',\delta),
TN\log\left(\frac{TN}{\delta}\right),\frac{1}{\varepsilon_q}
\Bigg]
\right),
\end{align}
with the simplified $\log \Omega(\delta',\delta)$ being
\begin{align}
\log \Omega(\delta',\delta)=\mathcal{O}\left(
\log\Bigg[
\frac{TN}{\delta'}
\log\left(\frac{TN}{\delta}\right)
\Bigg]
\right).
\end{align}
These expressions are the starting point for the complexity analysis presented in Section~\ref{complexity_section}, where they are re-expressed in terms of the physically meaningful parameters $T$, $\varepsilon_c$, and $\varepsilon_q$.

\section{Classical complexity and error analysis}\label{appendix:classical_error_complexity}
To compare with the quantum algorithm, we estimate the computational complexity and discretization error of a straightforward classical implementation of the finite-difference scheme given in Eqs.~(\ref{eq:1species_discretized_a}-\ref{eq:1species_discretized_d}). To obtain the distribution matrix at time $T=mh$, i.e. for $f(T)$, one needs to perform  
\begin{align}
\mathcal{O}\left(km N \right)
\end{align}
elementary operations, where $k$ is the order of the time integrator, $m$ is the number of time-steps and $N=N_xN_v$ is the total number of grid-points. Note that the double integral is cumulative in $x$, hence it does not introduce any additional $N_x$ or $N_v$ factors. The local truncation error at each time-step is of the same asymptotic order as that introduced during the first step and is given by
\begin{align}
\varepsilon_c(h)=\mathcal{O}\left(h^{k+1}+h\left(\Delta x^2+\Delta v^2\right)\right),\label{eq:classical_error_h}
\end{align}
since the phase-space discretization is second order. After $m$ steps it accumulates to
\begin{align}
\varepsilon_c(T)=\mathcal{O}\left(m\left[h^{k+1}+h\left(\Delta x^2+\Delta v^2\right)\right]\right).\label{eq:eps(T)}
\end{align}
Now, typically stability conditions require $h\leq \mathcal{O}(\Delta x^q + \Delta v^q)$, where $q\geq 1$ is some low integer. $k$ is usually fixed in the range $[2,5]$. In these practical cases then the first contribution in Eq. \eqref{eq:eps(T)} is either asymptotically smaller than the second one or scales the same way. These two cases result in the same scaling in big $\mathcal{O}$ notation. Therefore, for a grid with a fixed $N_x$ and $N_v$ ratio, i.e with $N_x^2=\mathcal{O}(N)=N_v^2$, we have
\begin{align}
\varepsilon_c(T)\approx\mathcal{O}\left(\frac{T}{N}\right),
\end{align}
because $\Delta x = \mathcal{O}\left( 1/\sqrt{N} \right)=\Delta v$ and $m=T/h$. So the classical time complexity is approximately
\begin{align}
\approx\mathcal{O}\left(m \cdot \frac{T}{ \varepsilon_c}\right) = \mathcal{O}\left(T^{2} \varepsilon_c^{-1}\right),
\end{align}
where $\varepsilon_c=\varepsilon_c(T)$ is understood. This is now a function of $\varepsilon_c$ and $T$ only, hence mathematically comparable to the quantum gate complexity. $k$ can be ignored as it is fixed.

\end{document}